\documentclass[utf8]{frontiersFPHY} % for Physics and Applied Mathematics and Statistics articles

\usepackage{url,hyperref,lineno,microtype,subcaption}
\usepackage[onehalfspacing]{setspace}
\usepackage{subcaption}

\def\keyFont{\fontsize{8}{11}\helveticabold }
\def\firstAuthorLast{Ghosh {et~al.}} %use et al only if is more than 1 author
\def\Authors{Suprovo Ghosh\,$^{1}$, Bikram Keshari Pradhan\,$^{1}$, Debarati Chatterjee\,$^{1,*}$, J\"urgen Schaffner-Bielich\,$^{2}$}
\def\Address{$^{1}$ Inter-University Centre for Astronomy and Astrophysics, \\
Pune University Campus, Pune - 411007, India \\
$^{2}$ Institut f\"ur Theoretische Physik, Goethe Universit\"at, \\
Max von Laue-Stra\ss e 1, 60438 Frankfurt am Main, Germany  }
\def\corrAuthor{Corresponding Author}

\def\corrEmail{debarati@iucaa.in}

\begin{document}
\onecolumn
\firstpage{1}

%\title[Running Title]{Article Title} 
\title[Multi-physics constraints on hyperonic stars]{Multi-physics constraints at different densities to probe nuclear symmetry energy in hyperonic neutron stars} 
%\title[Hyperons and Symmetry energy in Neutron Stars]{Hyperons in Neutron Stars and Symmetry Energy in the Relativistic Mean-Field Model}

\author[\firstAuthorLast ]{\Authors} %This field will be automatically populated
\address{} %This field will be automatically populated
\correspondance{} %This field will be automatically populated

\extraAuth{}% If there are more than 1 corresponding author, comment this line and uncomment the next one.
%\extraAuth{corresponding Author2 \\ Laboratory X2, Institute X2, Department X2, Organization X2, Street X2, City X2 , State XX2 (only USA, Canada and Australia), Zip Code2, X2 Country X2, email2@uni2.edu}

\maketitle

\begin{abstract}
%The appearance of strangeness in the form of hyperons within the inner core of neutron stars is expected to affect its detectable properties such as its global structure or gravitational wave emission. The symmetry energy is a key quantity that governs the behaviour of dense nuclear matter. It would therefore be interesting to investigate how symmetry energy depends on nuclear and hypernuclear properties. In this work, we study neutron stars including hyperons within the framework of the Relativistic Mean Field model and explore the parameter space allowed by present uncertainties in nuclear and hypernuclear experimental data subject to multimessenger constraints. We investigate possible correlations between nuclear empirical parameters, particularly the symmetry energy, with observable properties of neutron stars.
%%% Leave the Abstract empty if your article does not require one, please see the Summary Table for full details.
%\section{}
%For full guidelines regarding your manuscript please refer to \href{http://www.frontiersin.org/about/AuthorGuidelines}{Author Guidelines}.

The appearance of strangeness in the form of hyperons within the inner core of neutron stars is expected to affect its detectable properties such as its global structure or gravitational wave emission. In this work, we explore the parameter space of hyperonic stars within the framework of the Relativistic Mean Field model allowed by present uncertainties in state-of-the-art nuclear and hypernuclear experimental data. We impose multi-physics constraints at different density regimes to restrict the parameter space: Chiral effective field theory, heavy-ion collision data as well as multi-messenger astrophysical observations of neutron stars. We investigate possible correlations between empirical nuclear and hypernuclear parameters, particularly the symmetry energy and its slope, with  observable properties of neutron stars. We do not find a correlation for the hyperon parameters and the astrophysical data. However, the inclusion of hyperons generates a tension between the astrophysical and heavy ion data constraining considerable the available parameter space.

\tiny
 \keyFont{ \section{Keywords:} symmetry energy, hyperons, neutron stars, equation of state, nuclear, hypernuclear, heavy-ion, multi-messenger}
  %All article types: you may provide up to 8 keywords; at least 5 are mandatory.
\end{abstract}

\textbf{Abstract length: 149 words\\
Manuscript length: 5435\\
Number of Figures: 13\\
Number of Tables: 1}

\section{Introduction}
\label{sec:intro}

%\dc{Different probes of the QCD phase diagram: nuclear experiments, HICs, NSs}
Understanding strong interaction among hadrons is one of the most intriguing topics in nuclear physics. Despite the recent progress in understanding the phase diagram of Quantum Chromodynamics (QCD), the theory of strong interactions~\cite{Baym2018,QCDBook}, we are still far from achieving a unified description of nuclear matter under extreme conditions of density and temperature. While terrestrial nuclear experiments probe densities close to nuclear saturation density ($n_0 \sim$ 0.16 fm$^{-3}$)~\cite{Lattimer2015,Gandolfi2015}, heavy-ion collision (HIC) experiments~\cite{FOPI,ASY_EOS} provide information about hot and dense matter at several times $n_0$. Recent progress in Lattice QCD~\cite{Inoue:2016qxt,Inoue:2019jme,Fabbietti2020} are also providing new constraints on the properties of matter at high temperature and low densities. Neutron stars, on the other hand, are astrophysical laboratories that provide us an opportunity to investigate ultra-high density (up to 10 times $n_0$) and low temperature regime of the QCD phase diagram, given the conditions that exist only in its interior~\cite{GlendenningBook,Lattimer2004}. 
\\

Strangeness adds a new dimension to the description of nuclear matter. The presence of strangeness has already been established in heavy-ion collisions (appearance of hyperons and kaons) or in finite nuclear systems (hypernuclei). It is also conjectured that strangeness containing matter, in the form of hyperons, kaons or even deconfined quark matter, can appear at the ultra-high densities that exist in the core of a neutron star (NS). The appearance of strangeness can have significant impact on NS composition, structure and observable astrophysical properties, such as its mass, radius, cooling or gravitational wave (GW) emission~\cite{OertelRMP}.
\\

%\dc{Theoretical description of dense nuclear matter: EoS, $\chi EFT$}
In order to connect the NS internal composition with its global properties, one requires an Equation of State (EoS)~\cite{Lattimer2012,Lattimer2015,Baym2018}. The theoretical description of NS matter therefore requires the construction of models of hadron-hadron interaction, using non-relativistic (such as Skyrme or Gogny interactions)~\cite{Stone2007,Vautherin1972,Dutra2012} or relativistic (Relativistic Mean Field or Dirac-Brueckner-Hartree-Fock methods)~\cite{Serot1991,Serot1997} techniques. In the case of microscopic models (such as Brueckner-Hartree-Fock method), the interactions are rigorously calculated from lowest order term to increasing order. But the poor knowledge of three-nucleon forces limits their applicability to reproduce real astrophysical data.  Phenomenological models are more successful, with the model parameters usually constrained at densities close to $n_0$ and low values of isospin (neutron-proton ratio), but the uncertainty increases at larger densities and asymmetries. The nuclear symmetry energy (the difference between the binding energies of symmetric nuclear matter and neutron matter) is a key quantity that governs the difference in the behaviour of infinite symmetric nuclear matter and neutron star matter. 
\\

%\dc{Multi-messenger observations of NSs: EM+GW, Recent constraints on NS EoS: $M_{max}$, GW, NICER.}
Neutron stars are particularly interesting, as they can be observed via electromagnetic (X-ray, $\gamma$-ray, radio waves) as well as gravitational waves, opening up a new era of multi-messenger astronomy. Electromagnetic multi-wavelength observations of NSs reveal a wealth of details about its global structure~\cite{Demorest,Antoniadis}. NS masses can be determined to high precision using post-Keplerian effects in NSs in binary~\cite{Thorsett}. Traditionally radius measurements from thermal emission suffered from large uncertainties~\cite{Ozel2010,Steiner2013,Guillot2013}, but the recently launched NICER (Neutron Star Interior Composition Explorer)~\cite{NICER,NICER0030_Miller,NICER0030_Riley,NICER0740_Miller,NICER0740_Riley} mission has improved NS estimates by exploiting a novel scheme of modulation profiles of pulses. Finally with the recent detection of GWs for the first time from NS-NS (GW170817)~\cite{Abbott2017}, NS-BH (GW200105 and GW200115)~\cite{Abbott2021} and GW190425 ~\cite{GW190425} systems by LIGO ~\cite{adLIGO} and Virgo ~\cite{AdVirgo},   GW astronomy is allowing us to probe directly the interior of NSs. The tidal deformation of NSs in the strong gravitational field of its binary companion measured during the inspiral phase of the merger  depends on the EoS, and therefore reveals information about its radius and interior composition~\cite{Abbott2018,Abbott2019,Most,Annala}. 
\\

%\dc{Recap papers on constraints by Suprovo, Sabrina Huth, others}
In the recent past, there have been several attempts to impose constraints on the NS EoS by using data from NS multi-messenger astrophysical observations within a statistical Bayesian scheme~\cite{pang2021,Coughlin2019,Biswas2021,biswas2021prex,biswas2021bayesian,Dietrich2020,O_Boyle_2020}. In such a scheme, the low density EOS constrained by theoretical and experimental nuclear physics is matched with parametrized high density EOSs satisfying gravitational wave and electromagnetic data ~\cite{Capano2020,Tews_2018,Tews_2019a,Tews_2019b,Gandolfi_2019}. Usually the EoSs are based on different parametrization schemes such as piecewise polytropes~\cite{Annala,Hebeler2013,Read2009,Gamba2019}, spectral representation~\cite{Fasano2019,Lindblom2018}, speed-of-sound parametrization~\cite{Tews_2018,Greif2019,Landry2020} or nuclear meta-modelling technique \cite{Gueven2020}. Only a few recent works used the RMF model \cite{Traversi2020} or hybrid (nuclear + piecewise polytope) parametrizations \cite{Biswas2021} to obtain posterior distributions of empirical parameters. Correlations among empirical nuclear parameters and some chosen NS observables have only recently been explored~\cite{Zhang2019,Xie2019,Carson2019,zimmerman2020measuring}. Although several of these works suggested probing the effect of the presence of hyperons, none of them consistently included hyperons within such a scheme. It therefore remains to be investigated whether one can restrict the parameter space of uncertainties associated with hyperons (hypernuclear potentials or hyperon couplings) or if they show any physical correlations with measurable properties such as nuclear saturation parameters or NS astrophysical observables.
\\

%\dc{Aim of this work}
In a recent work~\cite{Ghosh2022}, multi-physics constraints were imposed at different density regimes on the nuclear EoS using a ``cut-off scheme", and correlations of nuclear saturation parameters with astrophysical observables were investigated. Motivated by the Bayesian approach, the parameters of the realistic nuclear model were varied within their allowed uncertainties, compatible with state-of-the-art nuclear experimental data and the parameter space constrained using a combination of current best-known physical constraints at different density regimes: theoretical (chiral effective field theory) at low densities, multi-messenger (multi-wavelength electromagnetic as well as GW) astrophysical data at high densities and experimental (nuclear and heavy-ion collision) at intermediate densities~\cite{Dexheimer2021,Tsang2018} to restrict the parameter space of the nuclear model. Further, nuclear and heavy-ion collision experiments are isospin symmetric (same number of neutrons and protons), so studying NS matter provides information about the unknown nuclear symmetry energy. 
\\

In this article, we extend the above investigation to neutron star matter including strangeness, particularly hyperons. Within the framework of the RMF model, and allowing for a parameter space spanning current uncertainties in nuclear and hypernuclear physics, we impose multi-physics constraints in different density regimes from terrestrial nuclear/hypernuclear and multi-messenger astrophysical data. The aim of this study is to investigate possible correlations between empirical nuclear and hypernuclear parameters (particularly the symmetry energy and its slope) with NS astrophysical observables. 
\\

%\dc{Structure of this article}
The structure of the article is as follows: in Sec.~\ref{sec:descrip}, we describe the methods used to determine the composition of NS matter including hyperons in the framework of the RMF model. In Sec.~\ref{sec:constraints}, we impose constraints at different densities on the hyperonic EoS. In Sec.~\ref{sec:results}, we discuss the results of this investigation and in Sec.~\ref{sec:discussions} we discuss the implications of these findings.
\\

%%%%%=====================================
\section{Methods}
%\section{Description}
\label{sec:descrip}

%\dc{EoS : RMF model} \\
As discussed in  Sec.~\ref{sec:intro}, we calculate the beta equilibrated, charge-neutral NS EoS within the RMF framework. For our investigation, we consider the standard baryon octet as well as electrons and muons. The interaction  Lagrangian density ($\mathcal{L}$) considered in this work is given in Eq.~\ref{eqn:lagr} ~\cite{ChatterjeeNPA,Pradhan2021}. In this model, the baryon-baryon interaction is mediated by the exchange of scalar ($\sigma$), vector ($\omega$), isovector ($\rho$) mesons, and the strange baryon, i.e., hyperon-hyperon interactions are carried out by additional strange scalar ($\sigma^*$) and strange vector ($\phi$) mesons.
\begin{eqnarray}
     \mathcal{L} &=&\sum_B  \bar{\psi}_{_B}  \Big(i\gamma^{\mu}\partial_{\mu}-m_{_B}+g_{\sigma B}\sigma-g_{\omega B}\gamma_{\mu}\omega^{\mu}-g_{\rho B}\gamma_{\mu} \vec{I_B}.\vec{\rho}^{\mu}\Big)\psi_{_B} + \frac{1}{2}  (\partial_{\mu} \sigma \partial^{\mu}\sigma - m_{\sigma}^2 {\sigma}^2)-U_{\sigma} \nonumber \\
     && \ \ +\frac{1}{2}m_{\omega}^2 \omega_{\mu}\omega^{\mu}-\frac{1}{4} \omega_{\mu \nu}\omega^{\mu \nu} -\frac{1}{4}  (\vec{\rho}_{ \mu \nu}.\vec{\rho}^{\mu \nu}-2 m_{\rho}^2 \vec{\rho}_{\mu}.\vec{\rho}^{\mu})
    + \Lambda_{\omega} (g_{\rho N}^2  \vec{\rho}_{\mu} .\vec{\rho}^{\mu}) \  (g_{\omega N}^2 \omega_{\mu}\omega^{\mu})  \nonumber \\ 
     && \ \ +\sum_Y  \bar{\psi}_{_Y}  (g_{\sigma^* Y} \sigma^*-g_{\phi Y}\gamma_{\mu}\phi^{\mu})\psi_{_Y}+\frac{1}{2}m_{\phi}^2 \phi_{\mu}\phi^{\mu} -\frac{1}{4} \phi_{\mu \nu}\phi^{\mu \nu}  +\frac{1}{2}  (\partial_{\mu} \sigma^* \partial^{\mu}\sigma^* - m_{\sigma^*}^2 {\sigma^*}^2)\nonumber \\
    && \ \ +\sum_{\ell=\{e^- , \ \mu^-\}}  \bar{\psi}_{\ell}  (i\gamma^{\mu}\partial_{\mu}-m_{\ell}){\psi}_{\ell}
     \label{eqn:lagr}
\end{eqnarray}
where,
 \begin{eqnarray}
     U_{\sigma}&=&\frac{1}{3}b m_N   (g_{\sigma N} \sigma)^3+\frac{1}{4}c   (g_{\sigma N} \sigma)^4 \nonumber
\end{eqnarray}
In Eq. ~\ref{eqn:lagr}, $B$ stands for the baryon octet ($p,\ n,\ \Lambda,\ \Sigma^-,\ \Sigma^0,\ \Sigma^+,\ \Xi^-,\ \Xi^0$) whereas $Y$ stands for hyperons ($\Lambda,\ \Sigma^-,\ \Sigma^0,\ \Sigma^+,\ \Xi^-,\ \Xi^0$). One can solve the equation of motion   governing constituent particle  fields ($\Psi$) as well as those of mesons following ~\cite{Hornick,Pradhan2021}. Replacing the meson fields with their mean values in RMF framework, the energy density ($\epsilon$) corresponding to the Lagrangian given in Eq.~\ref{eqn:lagr} can be expressed as,
\begin{eqnarray}
    \epsilon &=&\frac{1}{2}m_{\sigma}^2\bar{\sigma}^2+\frac{1}{2}m_{\omega}^2\bar{\omega}^2+\frac{1}{2}m_{\rho}^2\bar{\rho}^2+\frac{1}{2}m_{\sigma^*}^2\bar{\sigma^*}^2  + \frac{1}{2}m_{\phi}^2\bar{\phi}^2+\frac{1}{3}b m_N   (g_{\sigma N} \bar{\sigma})^3+\frac{1}{4}c   (g_{\sigma N} \bar{\sigma})^4 \nonumber\\
    &+& 3 \Lambda_{\omega} (g_{\rho N} g_{\omega N} \bar{\rho} \bar{\omega} )^2 +\sum_B \frac{g_{sB}}{2\pi^2} \ \int_0^{k_{FB}} {\sqrt{k^2+{m_B^*}^2}}\  dk + \sum_{\ell} \frac{g_{s\ell}}{2\pi^2} \ \int_0^{k_{F{\ell}}} {\sqrt{k^2+{m_{\ell}}^2}}\  dk \label{eqn:endens}
\end{eqnarray}

Where $g_{si}$ and $k_{Fi}$ represent the spin degeneracy and Fermi momentum of the `$i^{th}$' species respectively. The baryon effective mass ( $m_B^*$) is then defined as $m_B^*=m_B-g_{\sigma B}\bar{\sigma}-g_{\sigma^* B}\bar{\sigma^*}$.  The pressure can be expressed using Gibbs-Duhem relation,
\begin{equation}
     p=\sum_{i=B,\ell} \mu_i n_i-\epsilon 
     \label{eqn:pres}
 \end{equation}
 with $n_i$ being the number density of $i^{th}$ constituent . The  chemical potentials for baryon ($\mu_B$) and lepton ($\mu_{\ell}$) are given by,
 \begin{eqnarray}
   \mu_B &=& \sqrt{k_{FB}^ 2  + {m^*_B}^2 }  + g_{\omega_B} \bar{\omega}+g_{\phi_B} \bar{\phi}+I_{3_B} g_{\rho_B} \bar{\rho}\, \nonumber \\
   \mu_{\ell}&=&\sqrt{k_{F\ell}^ 2  + {m_{\ell}}^2}\,.
    \label{eqn:chempot}
\end{eqnarray}

\subsection{Nucleonic matter}

The isoscalar nucleon-meson coupling parameters ($g_{\sigma N},\ g_{\omega N},\ b,\ \text{and }c$) are determined by fixing the nuclear saturation parameters: nuclear saturation density ($n_0$), binding energy per nucleon  at saturation ($E_{sat}$), incompressibility ($K_{sat}$) and the effective nucleon mass ($m^*$) at saturation. On the other hand the isovector couplings `$g_{\rho N}$' and `$\Lambda_{\omega}$' are fixed to the symmetry energy ($E_{sym}$) and slope of symmetry energy ($L_{sym}$) at saturation ~\cite{Hornick,Chen2014,Ghosh2022}. The range of empirical parameters considered in this work are consistent with state-of-the-art nuclear experimental data~\cite{Ghosh2022} and are summarised in Table.~\ref{tab:rangepara}.  \\

\iffalse
Nuclear parameter space (same as our paper): \\
$n_0$: 0.14-0.17 fm$^{-3}$ \\
$E_{sat}$: -16+-0.2 MeV \\
$K_{sat}$: 200-300 MeV \\
$E_{sym}$:  28-34 MeV \\
$L_{sym}$: 40-70 MeV \\
$m^*/m$: 0.55-0.75
\fi

\subsection{Hyperonic matter}
\label{sec:hyperon}
 In the Lagrangian (Eq.~\ref{eqn:lagr}), the attractive interaction among hyperons is meditated by the exchange of strange scalar ($\sigma^*$) meson and the repulsive interaction is mediated by exchange of strange vector ($\phi$) meson. However, it has been concluded that models with attractive hyperon-hyperon interaction show incompatibility with observations of the maximum NS mass~\cite{ChatterjeeNPA}. Hence we set the strange scalar couplings to 0, i.e, $g_{\sigma Y}=0$ and the remaining  non-strange hyperon meson coupling constants ($g_{\sigma Y}$) are fitted to the hyperon-nucleon potential ($U_Y$) at saturation using Eq.~\ref{eqn:hyppot} ~\cite{ChatterjeeNPA}. Among the nucleon-hyperon potentials, the best known potential is that of the hyperon $\Lambda$, having a value of $U_{\Lambda}=-30$ MeV ~\cite{Millener88,Schaffner92}. Although the potential depths for hyperons $\Sigma$ and $\Xi$ are not known precisely, it has been concluded that the $\Sigma$-nucleon potential is repulsive ~\cite{FriedmanGal07,SchaffnerGal00,Mares951} whereas $U_{\Xi}$ is attractive in nature ~\cite{Fukuda98,Khaustov00,SchaffnerGal00}. Hence for  this investigation we vary  $U_{\Sigma}$ in the range of 0 to +30 MeV\ and $U_{\Xi}$ from -30MeV to 0. The vector hyperon couplings ($g_{\omega Y},g_{\phi Y}$) are fixed to their SU(6) values (see Eq.~\ref{eqn:su6})~\cite{Schaffner93,ChatterjeeNPA}.
 
 \begin{equation}
    U_{Y} (n_0)=-g_{\sigma Y}\bar{\sigma}+g_{\omega Y}\bar{\omega}
    \label{eqn:hyppot}
\end{equation}
 \begin{eqnarray}
    g_{\omega {\Lambda}}=g_{\omega {\Sigma}}=2g_{\omega {\Xi}}&=&\frac{2}{3}g_{\omega N} \nonumber\\
  2g_{\phi {\Lambda}}= 2g_{\phi {\Sigma}}=g_{\phi {\Xi}}&=&\frac{-2\sqrt{2}}{3}g_{\omega N}
    \label{eqn:su6}
\end{eqnarray}

\begin{figure}[htbp]
\centering
\begin{minipage}[c]{0.5\textwidth}
  \centering
  \includegraphics[width=1\linewidth,height=0.78\linewidth, angle=0]{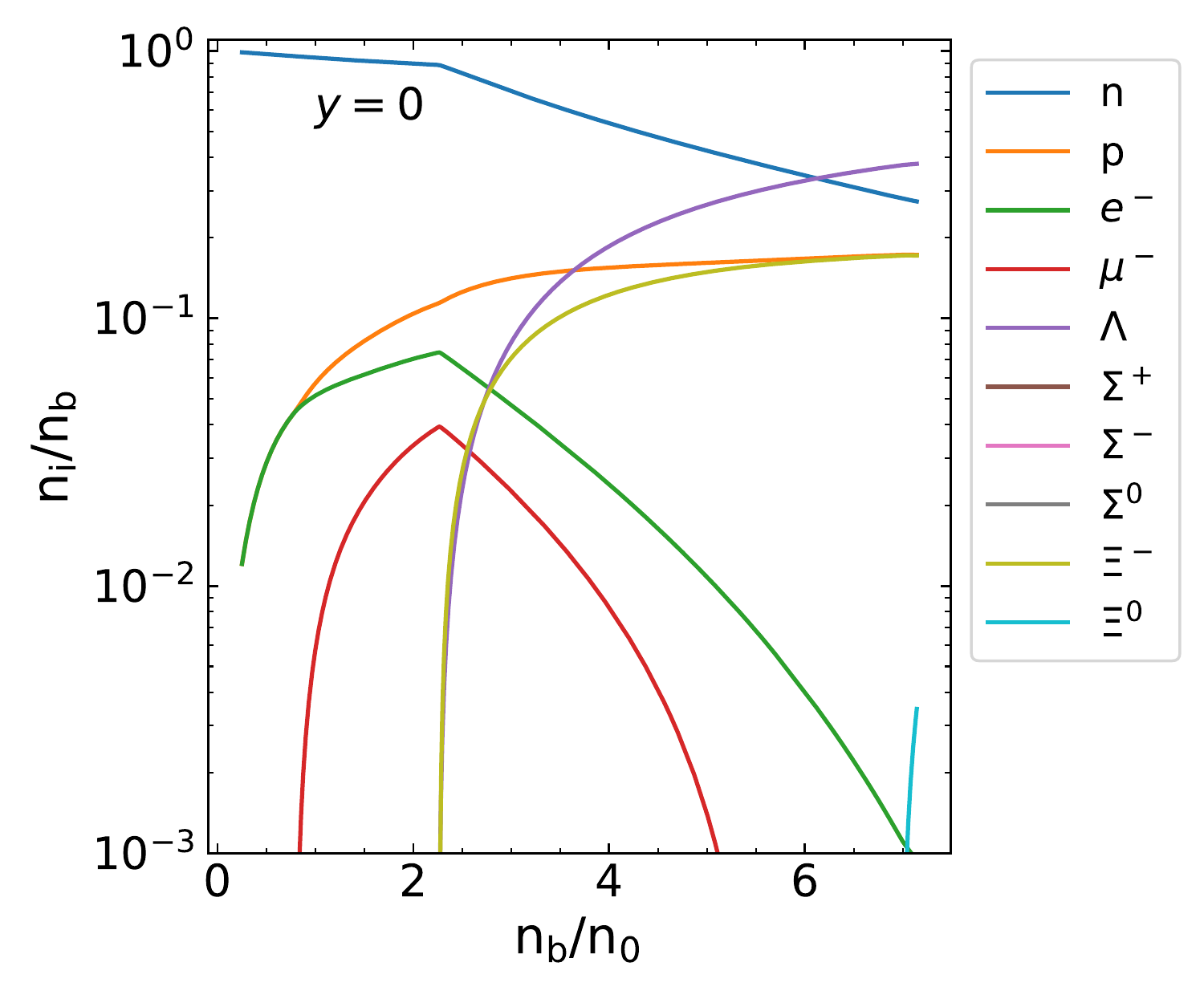}
  \text{\textbf{(A)} Particle fractions with $y$ = 0 }
\end{minipage}
\begin{minipage}[c]{0.5\textwidth}
  \centering
  \includegraphics[width=1\linewidth,height=0.78\linewidth, angle=0]{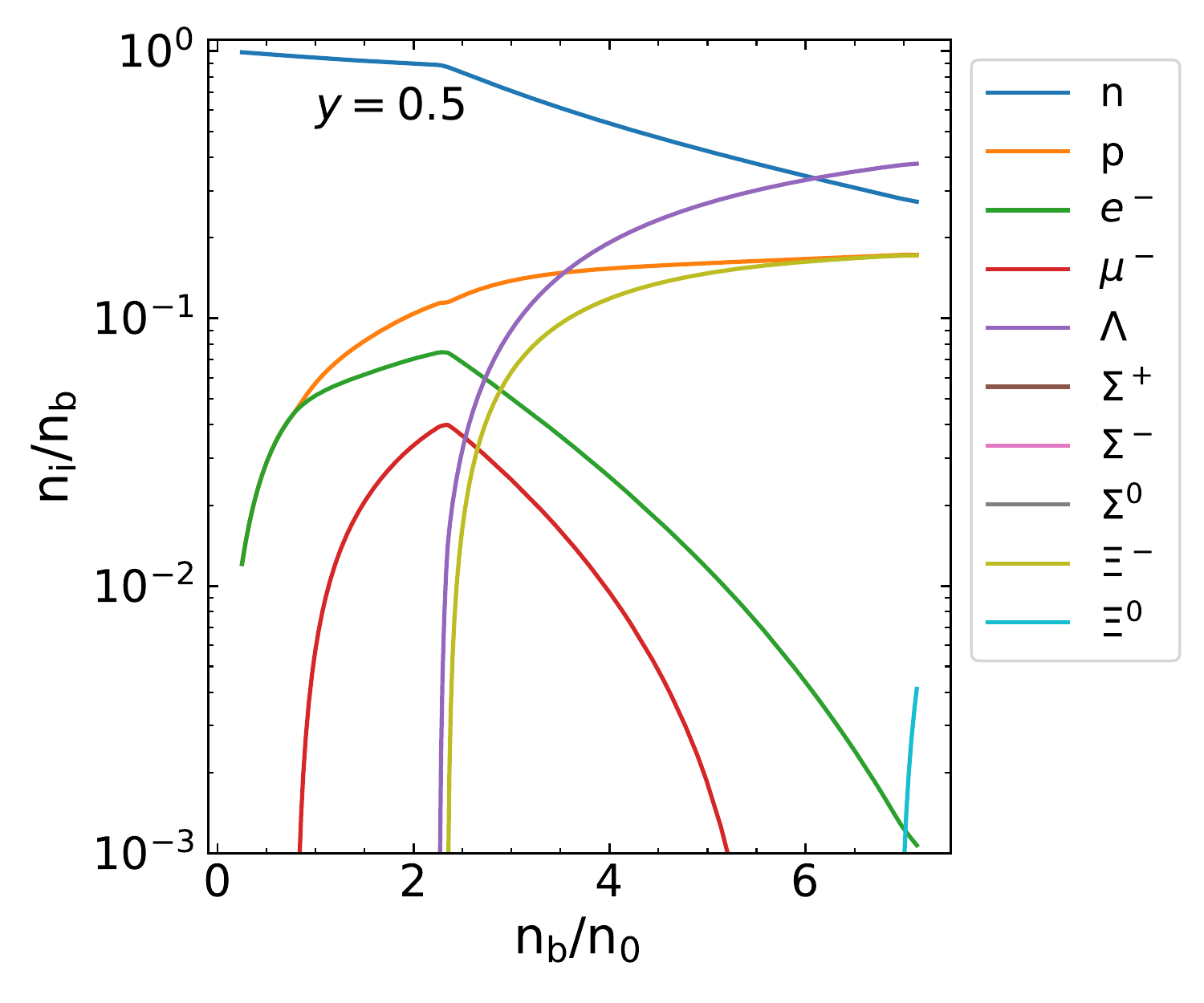}
  \text{\textbf{(B)}  Particle fractions with $y$ = 0.5 }
\end{minipage}
\begin{minipage}[c]{0.5\textwidth}
  \centering
  \includegraphics[width=1\linewidth,height=0.78\linewidth, angle=0]{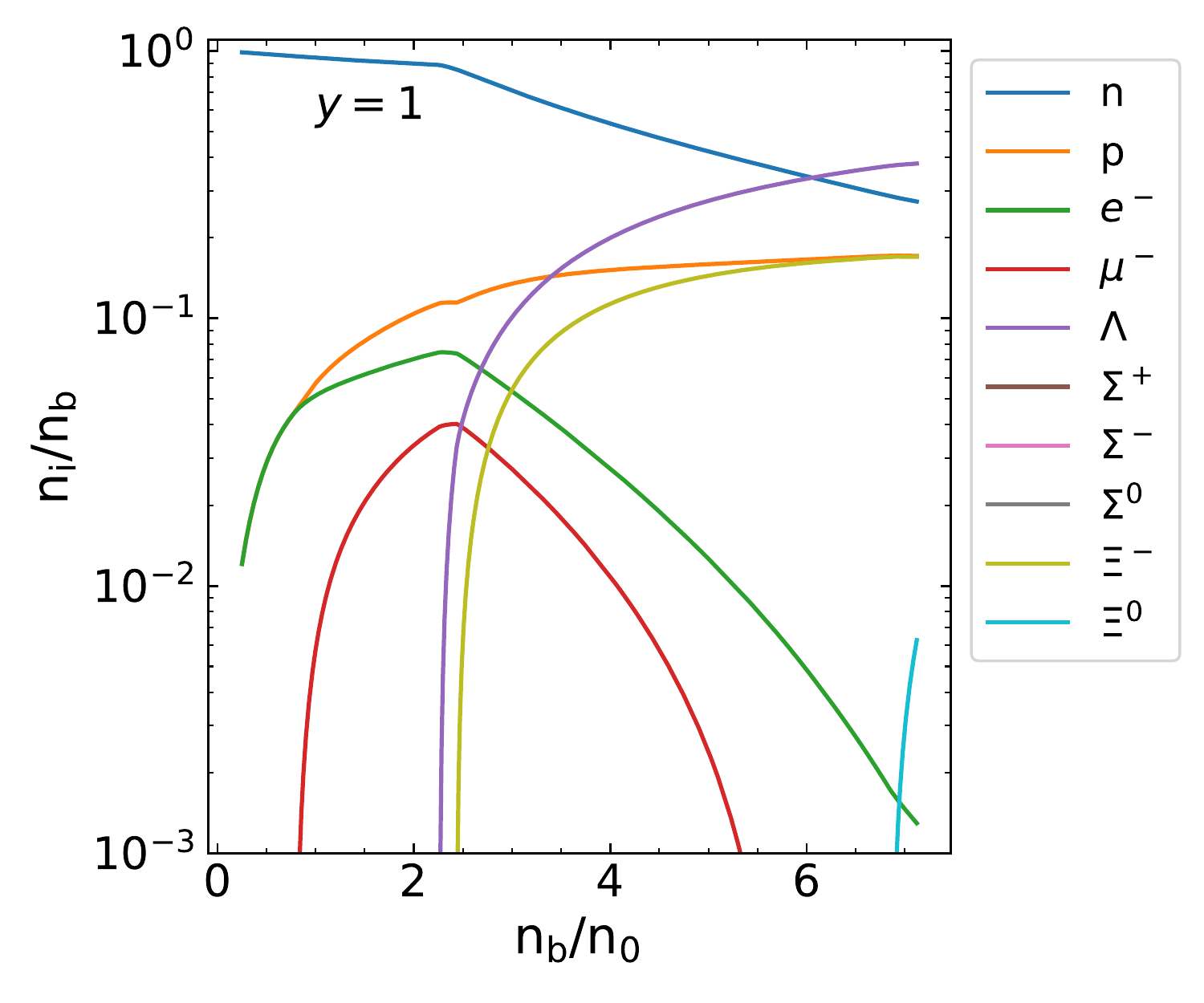}
  \text{\textbf{(C)}  Particle fractions with $y$ given by SU(6) symmetry }
\end{minipage}
\caption{Particle fractions for varying $y$ parameter (isovector hyperon coupling). The nuclear and hyper nuclear parameters are fixed to, $m^*=0.65m_N,\ E_{sat}=-16\  \rm{MeV},\  K_{{sat}}=240\  \rm{MeV},\ J_{{sym}}=32\ \rm{MeV},\ L_{sym}=60\  \rm{MeV}, \ U_{\Lambda}=-30\  \rm{MeV},\ U_{\Sigma}=+30 \ \rm{MeV},\ U_{\Xi}=-28\  \rm{MeV}.$ }
\label{fig:particlefrac_yparam}
\end{figure}

Recent lattice QCD calculations by the HALQCD group extracted hyperon potentials at almost physical quark masses and used it to estimate the hyperon potentials in pure neutron matter and in nuclear matter by using the Brueckner-Hartree-Fock approximation ~\cite{Inoue:2016qxt,Inoue:2019jme}. They find that the hyperon potentials in nuclear matter are $U_{\Lambda} = -28\ \rm{MeV},\  U_{\Sigma} = +15 \rm{MeV} and \ U_{\Xi} = -4 \ \rm{MeV}$. From their figure 4 in ~\cite{Inoue:2019jme} one can read off the hyperon potentials for $\Sigma^-$ and $\Sigma^+$, as well as for $\Xi^0$ and $\Xi^-$ in pure neutron matter. The relevant for neutron star matter are the ones for the $\Sigma^-$ and the $\Xi^-$ with the potentials of +25 MeV and + 6 MeV, respectively.
However, these results suggest that the isovector hyperon coupling ($g_{\rho Y}$) differs from its SU(6) quark model values. Where the isospin potential for the $\Xi$ would be as large as the one for nucleons and the one for the $\Sigma$ even twice as large. For a typical nucleon isospin potential of about 32 MeV one arrives at hyperon isospin potentials which are more than a factor two larger than the ones from the HALQCD analysis. We therefore introduce a scaling parameter `$y$,' which ranges from 0 to 1 (0 to SU(6) coupling strength) to span the uncertainty in the hyperon-isovector coupling. In that case, $g_{\rho Y}$ can be expressed as, 
%Recent simulation results coming from consideration of hyperons in Lattice QCD (HALQCD) suggest that the isovector hyperon coupling ($g_{\rho Y}$) differs from its SU(6) quark model values \dc{(References?)}. We therefore introduce a scaling parameter `$y$,' which ranges from 0 to 1 (0 to SU(6) coupling strength ) to span the uncertainty in the hyperon-isovector coupling. In that case, $g_{\rho Y}$ can be expressed as,
\begin{eqnarray}
g_{\rho \Lambda}&=&0 \nonumber \\
\frac{ g_{\rho {\Xi}}}{g_{\rho {N}} }&=& \frac{1}{2} \frac{g_{\rho {\Sigma}}}{ g_{\rho N}} =y, \  \  y \in [0,1].
\end{eqnarray}\label{eqn:isohyp}
The effect of the variation in the $y$-parameter (isovector hyperon coupling) on the particle fractions can be understood from Fig.~\ref{fig:particlefrac_yparam}. For a chosen parameter set from Table~\ref{tab:rangepara}, $\Xi^-$ starts to appear at 2.27, 2.34 and 2.45 $n_0$ for $y$ values 0, 0.5 and 1 respectively. However $\Xi^0$ starts to appear 6.977, 6.943 and 6.850 $n_0$ for $y$ changing from 0 to 1 in steps of 0.5. One must note that the threshold of appearance of the hyperons will depend upon the hyperon potentials and the nuclear saturation parameters chosen for the EoS.
\\

\iffalse
Hyperon potentials (as in the Weissenborn NPA paper) :\\
$U_{\Lambda}$ : -30 MeV (fixed) \\
$U_{\Xi}$ : -40 - +40 MeV \\
$U_{\Sigma}$ : -40 - +40 MeV \\
\fi

\iffalse
\subsubsection{Suggestions}
\begin{enumerate}
    \item  variation of $U_{\Sigma}$ between 0 and +40 MeV (since $\Sigma$ potential is known to be repulsive) 
\item variation of hyperons isospin couplings from zero to its SU(6) value (motivation: the isospin coupling constants from HAL QCD are a factor 0.4 smaller compared to the SU(6) values) 
\item including the vector self-interaction term $\zeta$ (motivation: changes the high-density limit of the EOS. Currently under investigation by Bikram with project student Abhisek) 
\end{enumerate}
\fi

%%%%%%%%%%%%%%%%%%%%%%%%%%%%%%%%%%%%%%%%%%%%%%%%%%%%%%%%%%%%%%%%%%%%%%%%%%%%
\subsection{Global structure}
\label{sec:TOV}

%\dc{TOV: M, R, calculation of $\tilde{\Lambda}$}
The equilibrium structure of a non-rotating, relativistic NS is obtained by solving the coupled equations of hydrostatic equilibrium known as the Tolman-Oppenheimer-Volkof (TOV) equations~\cite{GlendenningBook,schaffner2020}
\begin{eqnarray}
\frac{dm(r)}{dr} &=& 4 \pi \varepsilon(r) r^2 ~, \nonumber \\
\frac{dp(r)}{dr} &=& - \frac{[p(r) + \varepsilon(r)] [m(r)+4 \pi r^3 p(r)] }{r(r-2 m(r))}~, \nonumber \\
%\frac{d \Phi(r)}{dr} &=& - \frac{1}{(\varepsilon + p)} \frac{dp}{dr}~.
\label{eq:tov}
\end{eqnarray}
With the given equation of state which gives a relation between the energy density ($\varepsilon$) and pressure ($p$), these TOV equations~\eqref{eq:tov} are integrated from the centre of the star to the surface with the boundary conditions of vanishing mass, $m|_{r=0}=0$, at the centre of the star, and a vanishing pressure, $p|_{r=R}=0$, at the surface. Varying the central density for a given EoS, we can get a sequence of NSs with different mass and radius, thus giving the M-R curves. \\

The tidal deformability parameter quantifies the degree of the tidal deformation effects due to the companion in coalescing binary NS systems during the early stages of an inspiral. It is defined as :
\begin{equation}\label{eq:tidal}
    \lambda = - \frac{Q_{ij}}{\varepsilon_{ij}}
\end{equation}
where $Q_{ij}$ is the induced mass quadrupole moment of the NS and $\varepsilon_{ij}$ is the gravitational tidal field of the companion. The tidal deformability $\lambda$ is related to the dimensionless $l = 2$ tidal Love number $k_2$ as~\cite{Hinderer,Hinderer2008}.
\begin{equation}
    \Lambda = \frac{2}{3} k_2 \left( \frac{R}{M} \right)^5~,
\label{eq:love}
\end{equation}
The tidal Love number ($k_2$) can be obtained by solving a set of differential equations coupled with the TOV equations~\cite{Yagi2013}. The total tidal effect of two neutron stars in an inspiraling binary system is given by the mass-weighted (dimensionless) tidal deformability $\tilde{\Lambda}$ defined as~\cite{Hinderer2010}
\begin{equation}\label{eq:dimtidal}
    \tilde{\Lambda} = \frac{16}{13}\frac{(M_1 + 12M_2)M_1^4\Lambda_1 + (M_2 + 12M_1)M_2^4\Lambda_2}{(M_1 + M_2)^5}
\end{equation}
where $\Lambda_1 = \Lambda_1(M_1)$ and $\Lambda_2 = \Lambda_2(M_2)$ are the (dimensionless) tidal deformabilities and $M_1, M_2$ are the masses of the individual binary components respectively. 
%\dc{M-R and M-$\tilde{\Lambda}$ figures}
\\

% For Original Research articles, please note that the Material and Methods section can be placed in any of the following ways: before Results, before Discussion or after Discussion.

%%%%%=====================================
\section{Multi-density constraints}
\label{sec:constraints}

%\dc{Recap of cut-off scheme from Suprovo's work} \\
In this work, we constrain the parameter space of the nucleonic and hyperonic matter as described in Sec.~\ref{sec:descrip} using a ``cut-off filter" scheme where we impose strict limits from nuclear and astrophysical observation to obtain the posteriors. In the language of Bayesian analysis, the priors are obtained by varying the nuclear empirical parameters,hyperon potentials and isovector couplings uniformly within their uncertainty range in Table~\ref{tab:rangepara} and the likelihood functions are appropriately chosen physical conditions as Filter functions described in Sec.~\ref{sec:filters}. \\

In Miller~et~al. (2019)~\cite{2019Miller}, they pointed out the statistical uncertainties in constraining EoS by putting strict limits from multi-messenger observations of neutron stars like we have used here. With a very large number of priors, this cut-off scheme gives a correct estimate of the nuclear parameter ranges consistent with the observations. Recent works~\cite{Annala,Most,2020Annala,annala2021multimessenger,refId0} also used similar cut-off schemes for constraining the EoS of ultra-dense matter. In the recent work by Ghosh~et~al.~\cite{Ghosh2022}, it was explicitly shown that including the statistical re-weighting using $\chi$-squared statistics might change the posterior probability distribution slightly, but it does not significantly alter the physical correlation between nuclear empirical parameters and astrophysical observables. So, we adopt this ``cut-off filter" scheme for this work. 

\subsection{Filter functions}
\label{sec:filters}
The following physical constraints  at different densities from multi-disciplinary physics are applied in this work - 
\begin{itemize}
\item \textbf{at low densities: $\chi EFT$}\\
Chiral EFT is an effective theory of QCD that describes strong many-body interactions among nucleons using order by order expansions in terms of contact interactions and long-range pion exchange interactions. In particular, the $\chi$ EFT expansion gives estimates of theoretical uncertainties depending on local chiral two and three-nucleon interactions using quantum Monte Carlo methods, which are one of the most precise many-body methods for nuclear physics~\cite{Drischler,Drischler2020}.The EoS of pure neutron matter (PNM) can be well constrained at low baryon densities $n_b$ in the range of $\sim$ 0.5-1.4 $n_0$~\cite{Drischler2019}. \\

\item \textbf{at high densities: NS astrophysical data}\\
   The constraints on the EoS at high density come from multi-messenger astrophysical observations, such as high mass NS observations, GW measurement of tidal deformability from binary neutron star mergers as follows :  
   \begin{enumerate}
       \item From the recent observations of the heaviest known pulsar PSR J0740+6620, the maximum mass of the neutron stars should be equal to or exceed $2.08^{+0.07}_{-0.07}$ $M_{\odot}$~\cite{fonseca2021refined}. This sets an upper bound on the maximum NS masses corresponding to the EoSs considered. 
       \item The recent analyses of the GW170817 event~\cite{Abbott2019} apply a constraint on the upper bound of the effective tidal deformability $\tilde{\Lambda} <$ 720~\cite{Tong2020} using the low-spin highest posterior density interval for tidal deformability. We do not consider the lower limit on tidal deformability in this study.  As explained in Sec.~\ref{sec:TOV}, the tidal deformability depends on the mass and radius (see Eq.~\ref{eq:love}), and therefore this result also leads to a constraint on the mass-radius relation~\cite{Most,Annala}.\\
        \end{enumerate}

\item \textbf{at intermediate densities: heavy ion collision experiments}\\
Heavy-ion collision experiments can provide additional information about the behaviour of hot dense matter at intermediate densities $\sim 1-3 n_0$. As in our previous investigation~\cite{Ghosh2022}, we impose constraints from three different heavy-ion collision experiments:
\begin{enumerate}
    \item \textbf{KaoS experiment :} Subthreshold $K^+$ meson production in Au+Au \& C+C nuclear collisions at the Kaon Spectrometer (KaoS) experiment at GSI, Darmstadt~\cite{Hartnack} yield kaon multiplicity, which is an indicator of the compressibility of dense matter at densities $\sim 2-3 n_0$. The analysis of the experimental data using Isospin Quantum Molecular Dynamics (IQMD) transport models points towards a soft EoS~\cite{Hartnack,Fuchs2001} and can be described by a simple Skyrme ansatz with an incompressibility $\lesssim$ 200 MeV. The constraint given by the KaoS data implies only those nucleon potentials which are more attractive than the  Skyrme  parametrization  within  the  considered  density regime will be allowed. 

     \item \textbf{FOPI experiment :} Elliptic flow data in Au+Au collisions between 0.4 and 1.5A GeV by the FOPI collaboration \cite{FOPI}, provides constraints for the EoS of compressed symmetric nuclear matter (SNM). Analysing the FOPI data using IQMD transport codes, one can obtain a constraint for the binding energy of SNM in the density region of $n_b/n_0 \sim 1.4-2.0$ ~\cite{FOPI}. To impose this constraint, the binding energy for SNM is calculated for the input parameters, and only permitted if they lie inside the band allowed by the FOPI data in this density range. 
     
     \item \textbf{ASY-EOS experiment :}  Information  about  the symmetry  energy for ANM at supra-saturation densities can be obtained from directed and elliptic flows of neutrons and light charged particles measured for the reaction $^{197}$Au+$^{197}$Au at 400 MeV/nucleon incident energy within the ASY-EOS experimental campaign at the GSI, Germany \cite{ASY_EOS}. To impose the ASY-EOS filter, the symmetry energy of ANM EoS is calculated for  the  input  parameters and allowed only if the symmetry energy lies inside the band allowed by the data in the range of $\sim 1.1 - 2.0 n_0$.
\end{enumerate}

\end{itemize}

\subsection{Correlations}
\label{sec:corrtheo}
Using the posterior obtained from the analysis, we look for any physical correlation of the nuclear parameters, hyperon potentials, isovector coupling among themselves and also with the astrophysical observables such as the mass and radius of the canonical 1.4$M_{\odot}$ and the massive 2$M_{\odot}$ NS. For this study, we use Pearson’s linear correlation coefficient defined as~\cite{ref1}.
\begin{equation}\label{weightcorrcoff}
    R_{XY} = \frac{Cov(X,Y)}{\sqrt{Cov(X,X)}\sqrt{Cov(Y,Y)}}
\end{equation}
where $Cov(X,Y)$ is the co-variance between two variables X and Y defined as
\begin{equation}\label{Cov}
 Cov(X,Y) = \frac{1}{N}\sum_i (X_i - M(X)) (Y_i - M(Y))
\end{equation}
where N is the no. of sample points and $M(X)$ is the mean of the variable X defined as $M(X) = \frac{1}{N}\sum_i X_i$.
%%%%%=====================================
\section{Results}
\label{sec:results}

\subsection{Effect of $\chi$EFT + astro filters}
\begin{figure}[htbp]
  \begin{center}
      \resizebox{0.7\textwidth}{!}{%
{\includegraphics{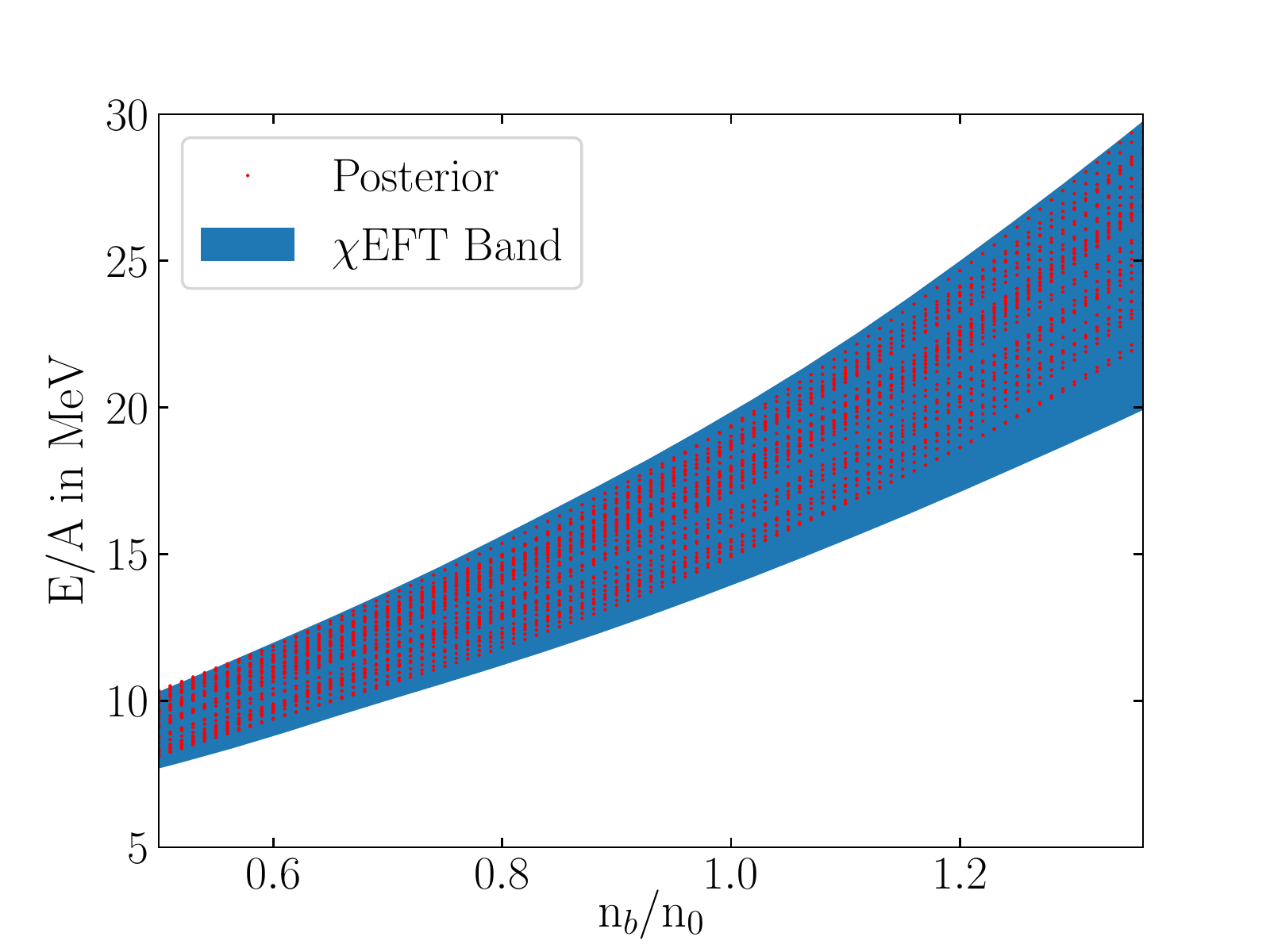}}
}
      \caption{Binding energy $E/A$ of posterior pure neutron matter EoSs as a function of normalized density $n_b/n_0$ allowed by chiral effective field theory ($\chi$EFT) data \cite{Drischler2019}.} 
    \label{fig:testposterior_binden} 
  \end{center}
\end{figure}

By randomly varying the parameter space of the nuclear parameters and hyperon potentials from Table.~\ref{tab:rangepara}, we generate the uniformly distributed prior set. After generating the random EoSs, we use the $\chi$EFT and astrophysical filters described in Sec.~\ref{sec:filters} to obtain filtered sets for the  parameters and NS observables.  For $\chi$EFT, we evaluate the binding energies in the density range of $n_b/n_0  \sim 0.5-1.4 $ corresponding to the $\chi$EFT data and allow only those parameter sets that lie within the band allowed by $\chi$EFT  calculations (Fig.~\ref{fig:testposterior_binden}).

\begin{figure}[htbp]
\centering
\begin{minipage}[b]{.7\textwidth}
  \centering
  \includegraphics[width=1.\linewidth, angle=0]{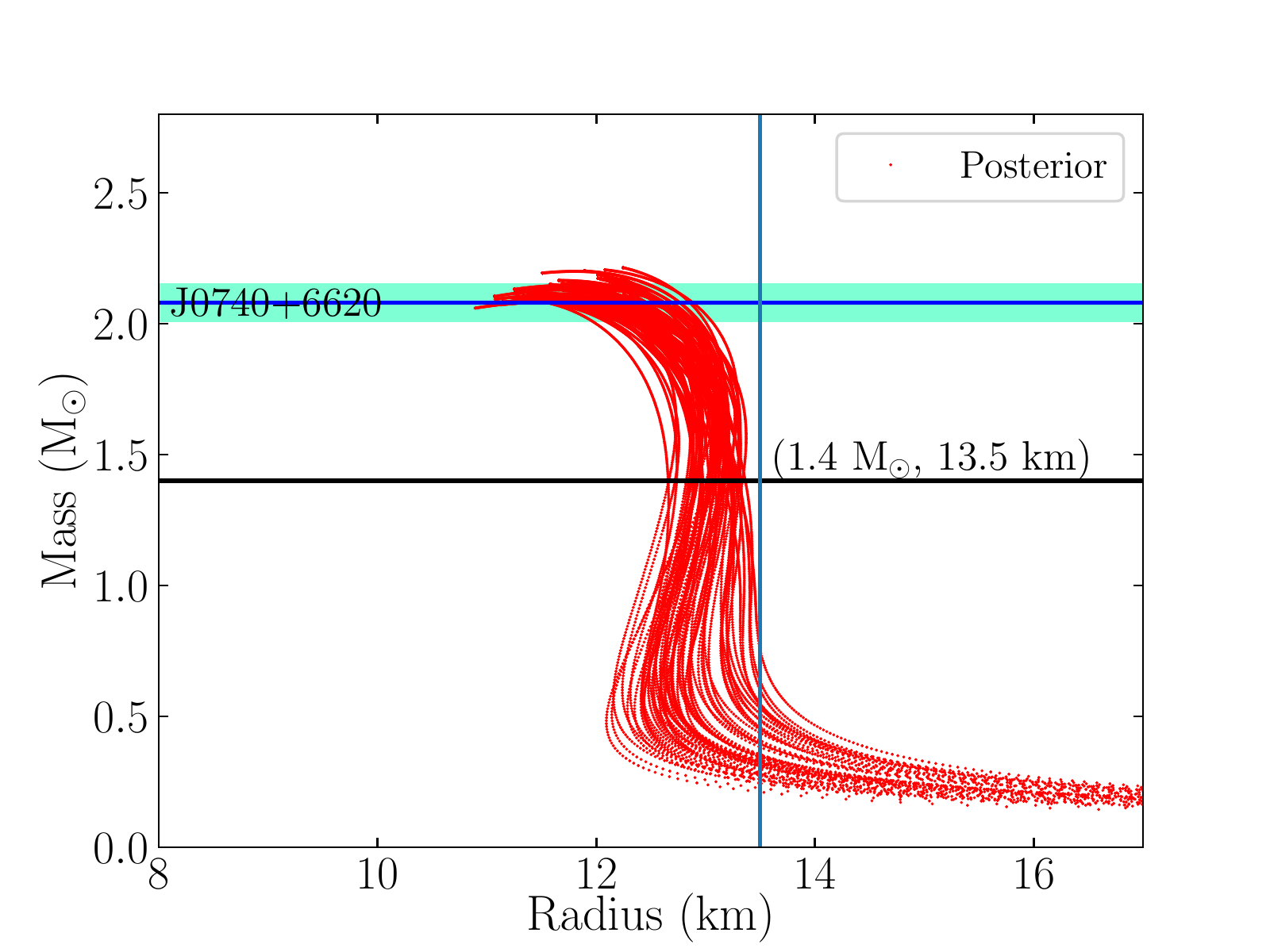}
  \text{\textbf{(A)} Mass-radius relation}
\end{minipage}
\begin{minipage}[b]{.7\textwidth}
  \centering
  \includegraphics[width=1\linewidth, angle=0]{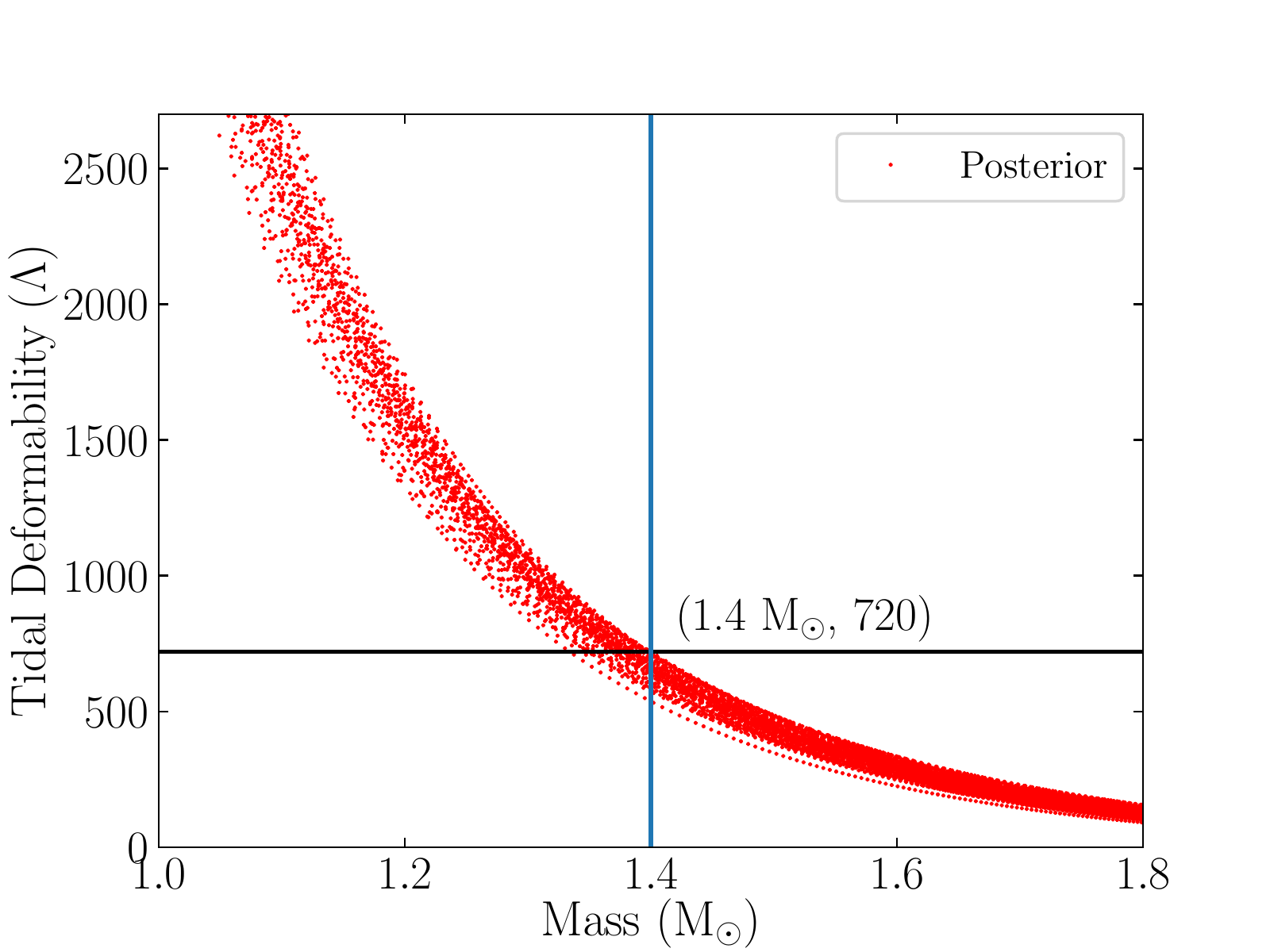}
  \text{\textbf{(B)} Dimensionless tidal deformability}
\end{minipage}
\caption{NS observables for the posterior hyperon EoSs after passing through $\chi$EFT and NS observations filters. The light green band indicates the uncertainty in the measurement of maximum mass of PSR J0740+6620~\cite{fonseca2021refined}}
\label{fig:posteriorastro}
\end{figure}

After obtaining the posterior $\chi$EFT, we use the same parameter set to obtain the hyperonic EoS using hyperon potentials and couplings given in Sec.~\ref{sec:hyperon}.We then solve the coupled TOV equations~\eqref{eq:tov}\eqref{eq:love} to obtain the mass, radius and tidal deformability of the NSs. Using the  multi-messenger astrophysical and GW observation of NSs given in Sec.~\ref{sec:filters}, we rule out further combinations of parameter sets and allow only those combinations which simultaneously satisfy all constraints on NS observables. In Fig.~\ref{fig:posteriorastro}, we plot the mass-radius relations and the dimensionless tidal deformability as a function of NS mass corresponding to the  filtered hyperonic EoSs. We can see that NS radii span a wide range from 11-14 km.

%There are 5 heading levels

\subsection{Correlations}
\label{sec:corr}
After obtaining the posterior parameter space, we look for any physical correlation among the parameters and the NS observables as well as within themselves.  In Fig.~\ref{fig:correlation}, we display the correlation matrix of the following quantities: nuclear empirical parameters ($n_0$, $E_{sat}$, $K_{sat}$, $E_{sym}$, $L_{sym}$), the effective mass $m^*/m$, hyperon potentials ($U_{\Sigma}$, $U_{\Xi}$),  hyperon-isovector coupling parameter $y$ and the NS observables ($R_{1.4M_{\odot}}$, $\Lambda_{1.4M_{\odot}}$, $R_{2M_{\odot}}$, $\Lambda_{2M_{\odot}}$) after applying both $\chi$EFT and astrophysical observations filter. Some of the main observations from the correlation matrix are listed below:
\begin{itemize}
    \item $n_0$ and $m^*/m$ show a high correlation (0.71).
    \item  $n_0$ has a moderate correlation with the NS observables. The correlation is noticeable (0.54) for the constraints for 1.4$M_{\odot}$ NS but is negligible for the constraints for 2~$M_{\odot}$.
    \item Symmetry energy $E_{sym}$ and its slope $L_{sym}$ display a strong correlation (0.79) which only appears when we apply the $\chi$EFT filter. This correlation only comes from the 
    $\chi$EFT filter around saturation density which is in agreement with previous literature~\cite{Hornick,Ghosh2022}.
    \item We see a moderate correlation (0.44) between $L_{sym}$ and effective mass $m^*/m$ after applying the $\chi$EFT filter.
    \item The correlation of $m^*/m$ with the NS observables is pretty low ($\sim 0.1$ for 1.4$M_{\odot}$ and $\sim 0.3$ for 2.0$M_{\odot}$ stars) which is quite the opposite to the purely nucleonic case~\cite{Ghosh2022}.
    \item The correlation between slope of symmetry energy $L_{sym}$ and radius of 1.4M$_{\odot}$ NS is also lower (around 0.4). A correlation between $L_{sym}$ and $R_{1.4 M_{\odot}}$ has been reported in several articles in the literature \cite{Fattoyev,Alam,Zhu,Lim}, although recent articles find $R_{1.4 M_{\odot}}$ to be nearly independent of $L_{sym}$~\cite{Hornick,Ghosh2022}.
    \item All the NS observables (radius and dimensionless tidal deformability for 1.4~$M_{\odot}$ and 2~$M_{\odot}$ NS), as expected, show a strong correlation with each other (according to Eq.~\ref{eq:tidal}) although we find a moderate correlation with the observables between 1.4~$M_{\odot}$ and 2~$M_{\odot}$ NSs.
    \item We did not find any correlation between the hyperon potentials and the isovector coupling parameter $y$ with other nuclear parameters and the astrophysical observables.
\end{itemize}

To understand the correlations better, in Fig.~\ref{fig:corner} we plot the posterior distribution of the nuclear parameters ($n_0$, $E_{sym}$, $L_{sym}$ and $m^*/m$) which show significant correlations and the astrophysical observables ($R_{1.4M_{\odot}}$, $R_{2M_{\odot}}$, $\Lambda_{2M_{\odot}}$) after applying both the $\chi$EFT and the astrophysical constraints.
\\

From the corner plots, we see that after applying the $\chi$EFT filter, both the median values of symmetry energy  and its slope $L_{sym}$  shift towards a higher value compared to their prior range in Table.~\ref{tab:rangepara} which leads to the strong correlation between them. For the effective mass, we see that the peak is around 0.63 which is lower than we observed for  purely nucleonic matter~\cite{Hornick,Ghosh2022}. This is because when we include hyperon, the EoS become softer~\cite{Pradhan2021}; so to satisfy the astrophysical constraint of maximum mass above 2$M_{\odot}$ and tidal deformability, the posterior of $m^*/m$ shifts towards a lower value. Also, from Fig 5 in~\cite{Ghosh2022}, we know that $\chi$EFT filter removes parameter sets with low effective mass and slope of symmetry energy which gives rise to a moderate correlation between $L_{sym}$ and $m^*/m$ observed here. For this reason, for the hyperonic case along with $\chi$EFT filter, the range of effective mass becomes narrow and peaks towards a lower value (around 0.63) which indeed restricts the radius of 1.4$M_{\odot}$ star to 12.6-13.4 km. This is why we observe a very low correlation between $m^*/m$ and the astrophysical observables. We also conclude that there is no correlation between the hyperon potentials ($U_{\Sigma},U_{\Xi}$) and the NS astrophysical observables.

\subsection{Effect of all filters : $\chi$EFT + astro + HIC (KaoS+FOPI+ASY-EOS)}
We first generated 50,000 prior sets, and on applying all filters ($\chi$EFT, astrophysical data, HIC) obtained almost no posterior sets. In order to understand the effect of the HIC filters, we then passed the prior sets only through the KaOS, FOPI and ASY-EOS filters and plotted the posterior of each nuclear parameters (see Fig.~\ref{fig:dist_hic}). From the figures, we observe that in the $K_{sat}$ distribution, the values are restricted below 240 MeV after HIC filters, and this is the effect of the KaoS filter. In the $L_{sym}$ distribution, the values are restricted to $>$ 55 MeV after HIC filters, resulting in a decreased correlation between $L_{sym}$ and $E_{sym}$. Both these effects were observed in nucleonic case also and discussed in our previous paper~\cite{Ghosh2022}. The most interesting is the distribution of $m^*/m$. In Sec.\ref{sec:corr}, we noted that including hyperons shifts $m^*/m$ to lower value for astrophysical filters. When we apply HIC filters, we see $m^*/m$ values peak at a higher value around 0.70. The distributions intersect at the two tail ends of the Gaussian curves, giving a very narrow range with low probabilities. This explains why there are so few posterior points due to the combined filters.
\\

In order to obtain a correlation plot after applying all filters, we generated a very large number (2 million) of priors after restricting the prior range to $m^*/m$ to 0.64 - 0.7 and managed to obtain 200 posterior sets. The resulting correlation plot with this set is given in Fig.~\ref{fig:correlations_all}. We see that effects of the HIC filters on correlations are the same as in the nucleonic case~\cite{Ghosh2022}: decrease of $L_{sym}-E_{sym}$ correlation, increase of $n_0$ correlation with NS astrophysical observables and increase of $K_{sat}-m^*/m$ correlation. We checked that allowing for hyperons mean that hyperons appear in all the cases investigated; $\Lambda$ hyperons always appear close to 2$n_0$, while the threshold for appearance of $\Sigma$ and $\Xi$ hyperons depend on the value of the corresponding hyperon potentials. One may also note that for 1.4$M_{\odot}$ stars, the fraction of hyperons in the core is lower than in more massive 2$M_{\odot}$ stars.
\\

%%%%%=====================================
\section{Discussion}
\label{sec:discussions}

\subsection{Summary of present results}
The motivation of this study is to investigate any existing correlations between empirical nuclear and hypernuclear parameters (particularly the symmetry energy and its slope) and with NS multi-messenger astrophysical observables such as its mass, radius and tidal deformability. To this aim, we extended our previous investigation~\cite{Ghosh2022} from nucleonic to hyperonic matter in NSs, i.e. within the framework of the RMF model, we constrained the parameter space allowed by current uncertainties in nuclear and hypernuclear physics, by imposing multi-physics constraints at different density regimes: chiral effective field theory at low densities, astrophysical constraints at high densities and heavy-ion collision data at intermediate energies. \\

First, using the filtered EoSs satisfying constraints from both $\chi$EFT and astrophysical data, we searched for any physical correlation among the parameters and the NS observables as well as among themselves. 
We found that the effective nucleon mass $m^*/m$ and saturation nuclear density $n_0$ show strong correlation. We found $n_0$ to be moderately correlated with radius and tidal deformability of 1.4$M_{\odot}$ NSs, but weakly correlated with those of 2$M_{\odot}$ stars. The correlation of $m^*/m$ with the NS observables was found to be low, contrary to the purely nucleonic case. It is interesting to note that the symmetry energy $E_{sym}$ and its slope $L_{sym}$ showed significant correlation after imposing the $\chi$EFT filter. There is a non-negligible correlation of $m^*/m$ with $L_{sym}$. 
\\

On applying all filters from $\chi$EFT, astrophysical and heavy-ion data, we found that very few nuclear parameter sets are able to satisfy all constraints simultaneously. By monitoring the individual posterior distributions of the nuclear saturation parameters, we confirmed the existence of a ``tension" between the constraints from the first two filters with those of heavy-ion data. The values of $K_{sat}$ are restricted to below 240 MeV due to the KaoS constraint and $L_{sym}$ to values larger than 55 MeV, drastically reducing the available parameter space. Further, low values of $m^*/m$ are allowed by astrophysical filters, while heavy-ion data allows large values. The overall effect of applying the heavy-ion filters was found to be same as in the nucleonic case~\cite{Ghosh2022}: a decreased $L_{sym}-E_{sym}$ correlation, increased $n_0$ correlation with NS astrophysical observables and enhanced $K_{sat}-m^*/m$ correlation.
\\

\subsection{Comparison with prior research}
There are several contrasting results in the hyperonic case as compared with the nucleonic case~\cite{Ghosh2022}. Mainly we found a decreased correlation of $m^*/m$ with the NS observables, and an increased correlation of $n_0$ with $m^*/m$ and NS observables for 1.4$M_{\odot}$ NS. 
Radii and dimensionless tidal deformability (for 1.4$M_{\odot}$ and 2$M_{\odot}$), show a strong correlation with each other as expected. However we find a moderate correlation with observables of 1.4$M_{\odot}$ with 2$M_{\odot}$ stars. This is due to the reduced range of radii for hyperonic stars. We checked that the distribution of $m^*/m$ shifts to lower values (peak around 0.63) in posteriors for hyperons compared to nucleons which peak around 0.7, restricting values of $R_{1.4 M_{\odot}}$ to a reduced range $\gtrapprox$ 13 km. 
The correlation between the slope of symmetry energy $L_{sym}$ and radius of 1.4$M_{\odot}$ NS is also lower than in the nucleonic case. A correlation between $L_{sym}$ and $R_{1.4 M_{\odot}}$ was reported in several articles in the literature~\cite{Fattoyev,Alam,Zhu,Lim}, although recent articles find $R_{1.4 M_{\odot}}$  to be nearly independent of $L_{sym}$~~\cite{Hornick,Ghosh2022}. 
%Finally, no correlation was found between the hyperon potentials or the isovector coupling parameter $y$ with other nuclear parameters and the astrophysical observables.
Finally, the astrophysical observables studied in this work (mass, radius, tidal deformability) do not seem to provide correlations with hyperon potentials or the isovector coupling parameter $y$. However one must note that this is not generic for all astrophysical observables. One has to look for other observables, which are sensitive to the hyperon content in the NS interior, such as $r$-modes, cooling, thermal evolution etc (see e.g. \cite{VidanaEPJA}).
\\

In another recent work~\cite{Traversi2020}, a Bayesian inference of the NS EoS was performed within the RMF model using astrophysical and nuclear saturation data. Using a selected class of nucleonic models with five empirical parameters and exploring different types of priors, they reported that the EoSs with the largest evidence were the ones featuring a strong reduction
of the nucleon effective mass. However, the major drawback of this model was the omission of interaction terms ($\Lambda_{\omega}$ in our work) in the Lagrangian, due to which other saturation parameters such as symmetry energy or its slope were not included. A preliminary investigation of the effect of hyperons was also performed by switching on only the $\Lambda$ hyperon, with a fixed potential depth and coupling constants. However the effect of the other baryons of the octet and variation of the couplings as well as their correlations with other nuclear saturation parameters or NS observables was unexplored.
\\

Another recent study~\cite{Gueven2020} used Bayesian statistics to combine low density nuclear physics data, such as the ab-initio $\chi$EFT predictions and the isoscalar giant monopole resonance, with astrophysical NS data, within the ``metamodel" approach for the dense matter EoS. The posterior probability distribution functions were marginalized over several higher order nuclear empirical parameters ($L_{sym}$, $K_{sym}$, $Q_{sat}$, $Q_{sym}$), and observational quantities such as radius of 1.4$M_{\odot}$ NS. This study also explored correlations among $L_{sym}-K_{sym}$ and $K_{sat}-Q_{sym}$ parameters, and reported marked tension between astrophysical and nuclear physics constraints. Some other work~\cite{biswas2021bayesian} combined laboratory experiments and NS astrophysical observation using Bayesian statistics
along with the LIGO/Virgo and NICER observations within a hybrid nuclear+piecewise polytrope (PP) EoS parameterization. This work
reported a very weak correlation between $L_{sym}$ and $R_{1.4 M_{\odot}}$. 
Recently Huth~et~al.~\cite{huth2021} used a Bayesian inference technique to analyse the nuclear EoS and NS properties, combining data from heavy-ion collisions (FOPI~\cite{FOPI} and ASY-EOS~\cite{ASY_EOS} experiments, EoS constraint for symmetric nuclear matter \cite{Danielewicz}), microscopic $\chi EFT$ calculations and multi-messenger information from NICER and XMM Newton missions, as well as GW data and the associated kilonova AT2017gfo9. The study concluded that HIC constraints to be in excellent agreement with NICER observations. However, hyperons were  not considered in the above investigations.
\\

\subsection{Limitations and future directions}
\label{sec:future}
In this work, the correlations between nuclear and hypernuclear parameters and NS astrophysical observables have been explored within the framework of the Relativistic Mean Field model. Although the advantage of this realistic phenomenological model is that, unlike polytropic or parametrized EoSs, the results provide important understanding of the underlying nuclear physics, it however remains to be established whether such physical correlations are also found in other realistic EoS models in order to generalise the results of this investigation. It would be interesting for example to see whether the conclusions would still hold for a Lagrangian with density-dependent couplings. Such possibilities will be addressed in a forthcoming publication. We recall here that the constraints from heavy-ion data are  model-dependent and should therefore be treated on a different footing than astrophysical constraints and their implications on the results discussed with a word of caution.
\\

In future, improved measurements of hyperon potentials in hypernuclear experiments, such as GSI in Germany, JLAB in USA and
J-PARC in Japan~\cite{VidanaEPJA}, will reduce the uncertainties in the hyperon-nucleon and hyperon-hyperon coupling strengths. With the advent of multi-messenger astronomy, new upcoming observations of NS properties will also help to provide more stringent constraints on the dense matter EoS in NSs.

%%%%%%%%%%%%%%%%%%%%%%%%%%%%%%%%%%%%%%%%%%%%%%%%

%\section{Nomenclature}
%\subsection{Resource Identification Initiative}
%To take part in the Resource Identification Initiative, please use the corresponding catalog number and RRID in your current manuscript. For more information about the project and for steps on how to search for an RRID, please click \href{http://www.frontiersin.org/files/pdf/letter_to_author.pdf}{here}.

%\section{Additional Requirements}
%For additional requirements for specific article types and further information please refer to \href{http://www.frontiersin.org/about/AuthorGuidelines#AdditionalRequirements}{Author Guidelines}.

\section*{Conflict of Interest Statement}
%All financial, commercial or other relationships that might be perceived by the academic community as representing a potential conflict of interest must be disclosed. If no such relationship exists, authors will be asked to confirm the following statement: 
The authors declare that the research was conducted in the absence of any commercial or financial relationships that could be construed as a potential conflict of interest.

\section*{Author Contributions}
S.G. and B.-K.P. have contributed equally to this work and share first authorship. D.C. and J.S.-B. share senior authorship and contributed to the conception of the study. D.C. is the corresponding author of this article. Both S.G. and B.-K. P. performed the theoretical, numerical and statistical analysis. D.C. wrote the first draft, S.G. and B.-K.P. wrote sections of the manuscript. All authors contributed to manuscript revision, read, and approved the submitted version.

\section*{Acknowledgments}
D.C. is grateful to the hospitality of the Institut f\"ur theoretische Physik, J. W. Goethe Universit\"at Frankfurt, Germany, where this work was carried out within the collaborative project “Astrophysical constraints for hyperons in neutron stars”. S. G. B.K.P. and D.C. acknowledge usage of the IUCAA HPC computing facility for the numerical calculations.

\section*{Funding}
%This collaboration was funded by the project B09: Probing strong-interaction matter with gravitational waves from neutron-star mergers of Transregios TR-211 (Strong interaction matter under extreme conditions). 
This work was supported by the Deutsche Forschungsgemeinschaft (DFG, German Research Foundation) through the CRC-TR 211 'Strong-interaction matter under extreme conditions'– project number 315477589 – TRR 211.

%\section*{Supplemental Data}
% \href{http://home.frontiersin.org/about/author-guidelines#SupplementaryMaterial}{Supplementary Material} should be uploaded separately on submission, if there are Supplementary Figures, please include the caption in the same file as the figure. LaTeX Supplementary Material templates can be found in the Frontiers LaTeX folder.

%\section*{Data Availability Statement}
%The datasets [GENERATED/ANALYZED] for this study can be found in the [NAME OF REPOSITORY] [LINK].
% Please see the availability of data guidelines for more information, at https://www.frontiersin.org/about/author-guidelines#AvailabilityofData

%\bibliographystyle{frontiersinSCNS_ENG_HUMS} % for Science, Engineering and Humanities and Social Sciences articles, for Humanities and Social Sciences articles please include page numbers in the in-text citations
\bibliographystyle{frontiersinHLTH&FPHY} % for Health, Physics and Mathematics articles
\bibliography{test}

%%% Make sure to upload the bib file along with the tex file and PDF
%%% Please see the test.bib file for some examples of references
%%%%%%%%%%%%%%%%%%%%%%%%%%%%%%%%%%%%%%%%%%%%%%%%%%%%%%%%
\newpage
\section*{Tables}
\begin{table}[h]
    \centering
    \caption{Range (minimum and maximum values) of nuclear and hypernuclear  parameters at saturation density used in this work. Masses of mesons and the nucleon are fixed as $m_{\sigma}=550 \ \rm{ MeV}, \ m_{\omega}=783\  \rm{ MeV}, \ m_{\rho}=770\  \rm{MeV}, \ m_{\phi}=1020\  \rm{MeV} \text{ and } \ m_N=939\  \rm{MeV} $.\\}
\begin{tabular}
{|p{1.5cm}|p{1.5cm}|p{1.5cm}|p{1.3cm}|p{1.3cm}|p{1.5cm}|p{1.cm}|p{1.5cm}|p{1.5cm}|p{0.9cm}|}
\hline
  $n_0$ & $E_{\rm{sat}}$& $K_{\rm{sat}}$& $E_{\rm{sym}}$ & $L_{\rm{sym}}$&$m^*$&$U_{\Lambda}$&$U_{\Sigma}$&$U_{\Xi}$&$y$\\ 
  ($\rm{fm}^{-3}$ )& (MeV) & (MeV)&  (MeV) &  (MeV) & ($m_N$) &  (MeV) &  (MeV) &  (MeV) & \\
 \hline
\hline
{} & {} & {} & {} & {} & {} & {} & {} & {} & {} \\
%0.14-0.17&-16$\pm$0.2&200 -300&28-34&40 -70&0.55 -0.75&-30&0 - 30&-30 - 0&0 - 1\\
0.14&-16.2&200&28&40&0.55&-30&0&-30&0\\
0.17&-15.8&300&34&70&0.75&-30&30&0&1\\
\hline
\end{tabular}
\label{tab:rangepara}
\end{table}

%%%%%%%%%%%%%%%%%%%%%%%%%%%%%%%%%%%%%%%%%%%%%%%%%%%%%%%%

\section*{Figures}

\iffalse
\begin{figure}[htbp]
\centering
\begin{minipage}[c]{0.8\textwidth}
  \centering
  \includegraphics[width=1\linewidth,height=0.78\linewidth, angle=0]{Figures/Correlation_Astro_NY.pdf}
  \text{\textbf{(A)} Filters : Astro }
\end{minipage}
\begin{minipage}[c]{0.8\textwidth}
  \centering
  \includegraphics[width=1\linewidth,height=0.78\linewidth, angle=0]{Figures/Correlation_CEFT_Astro_NY.pdf}
  \text{\textbf{(B)} Filters : $\chi$EFT + Astro}
\end{minipage}
\caption{Posterior correlation matrix for variation of nuclear empirical parameters and NS observables, after application of the $\chi$EFT and NS observations filter}
\label{fig:correlation}
\end{figure}
\fi

\begin{figure}[htbp]
  \begin{center}
      \resizebox{1\textwidth}{!}{%
{\includegraphics{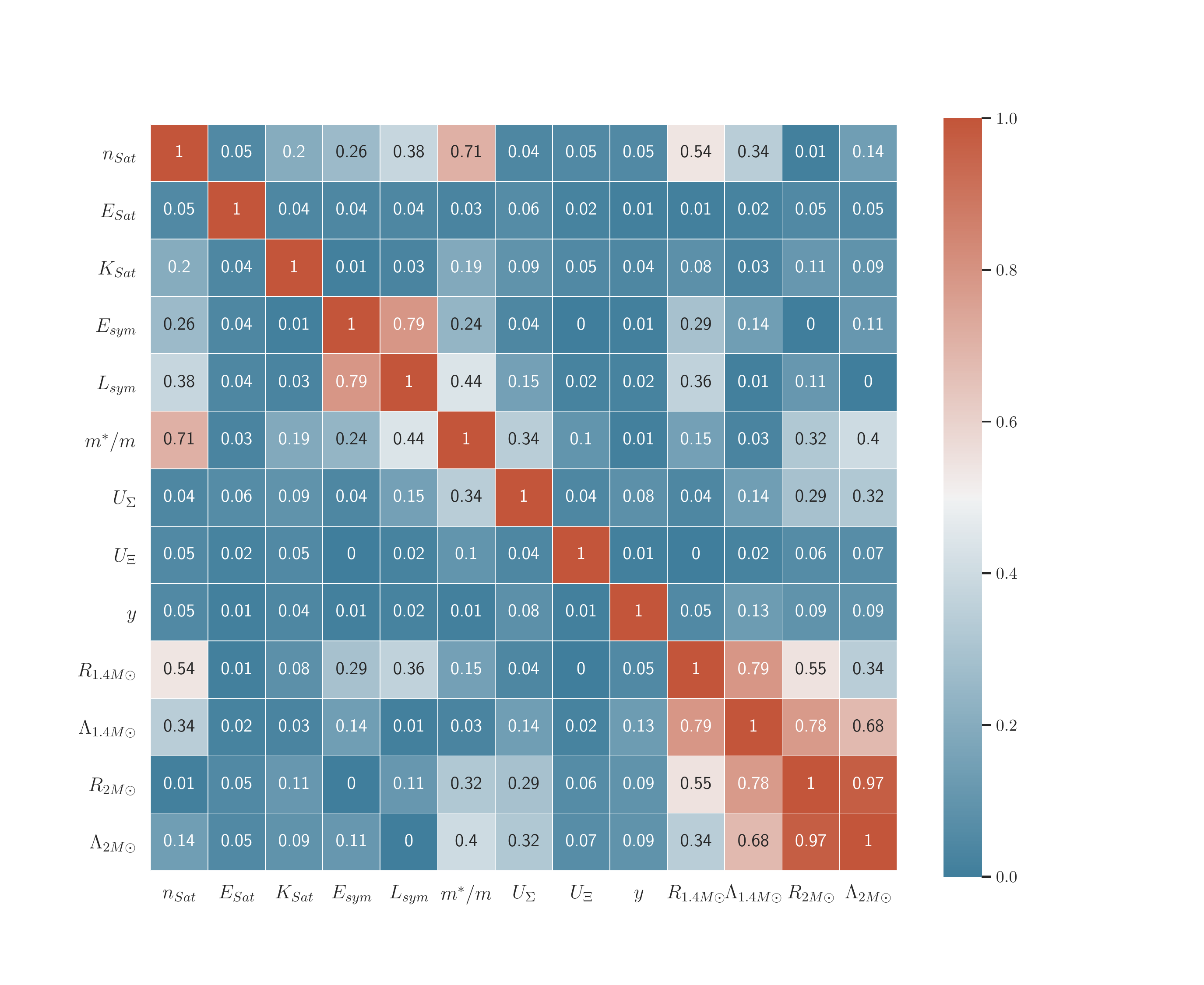}}
}
      \caption{Posterior correlation matrix for variation of nuclear empirical parameters and NS observables, after application of the $\chi$EFT and NS observations filter} 
    \label{fig:correlation} 
  \end{center}
\end{figure}

\begin{figure}[htbp]
  \begin{center}
      \resizebox{1\textwidth}{!}{%
{\includegraphics{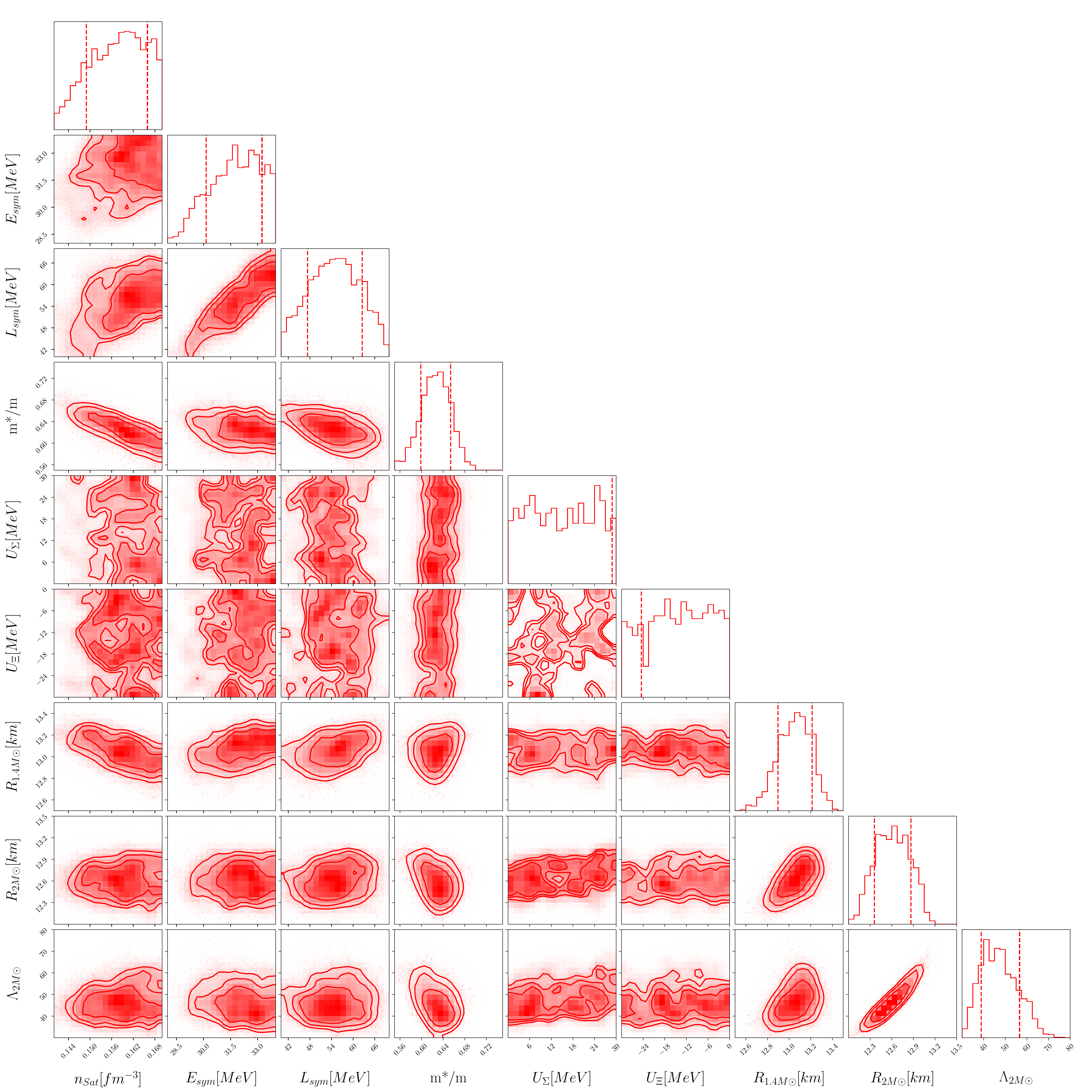}}
}
      \caption{Posterior distributions of nuclear parameters and astrophysical observables after applying the $\chi$EFT and the astrophysical constraints. } 
    \label{fig:corner} 
  \end{center}
\end{figure}

\begin{figure}[htbp]
\centering
\begin{minipage}[c]{0.45\textwidth}
  \centering
  \includegraphics[width=1\linewidth,height=0.78\linewidth, angle=0]{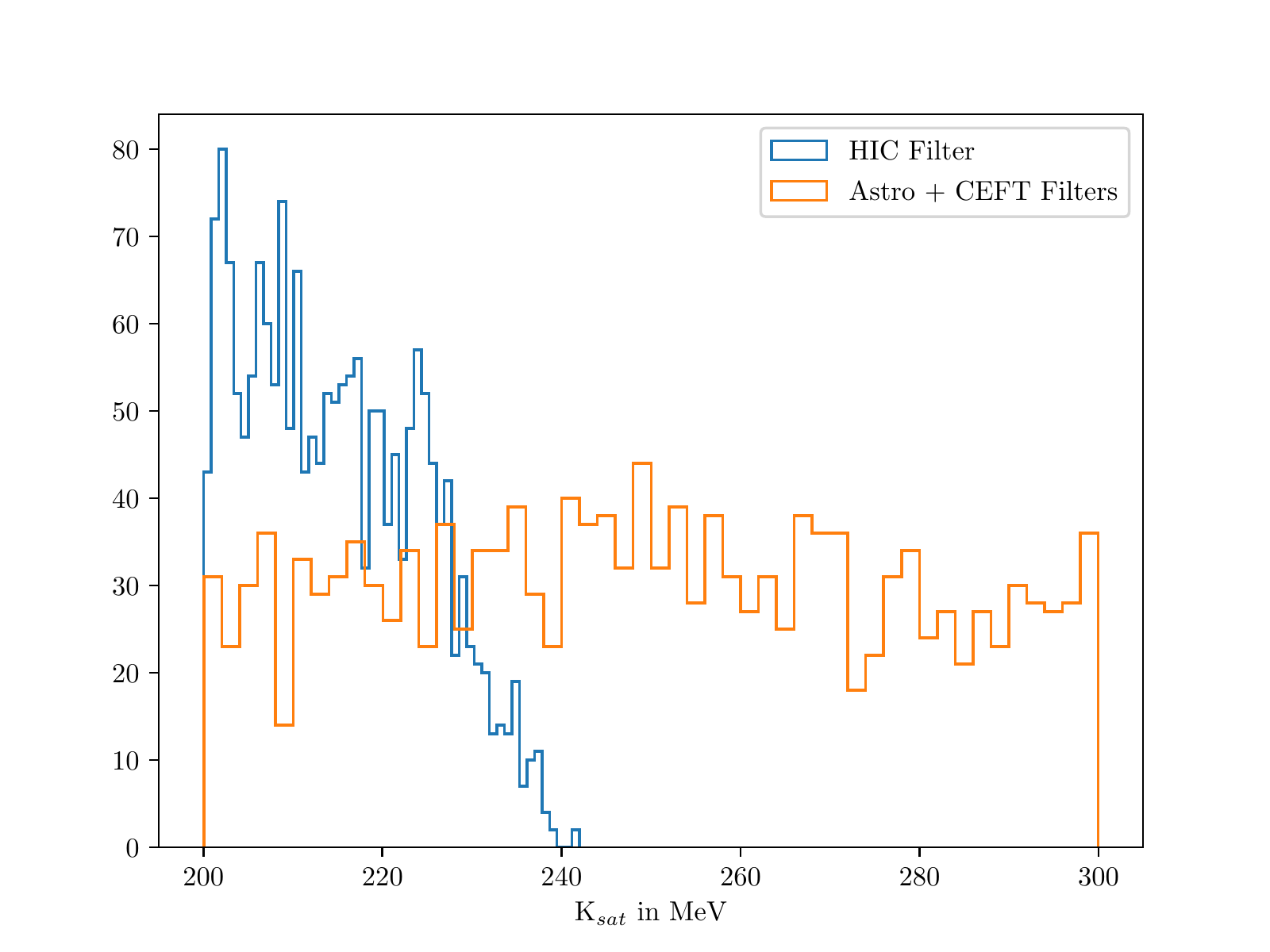}
  \text{\textbf{(A)} Distribution of $K_{sat}$ }
\end{minipage}
\begin{minipage}[c]{0.45\textwidth}
  \centering
  \includegraphics[width=1\linewidth,height=0.78\linewidth, angle=0]{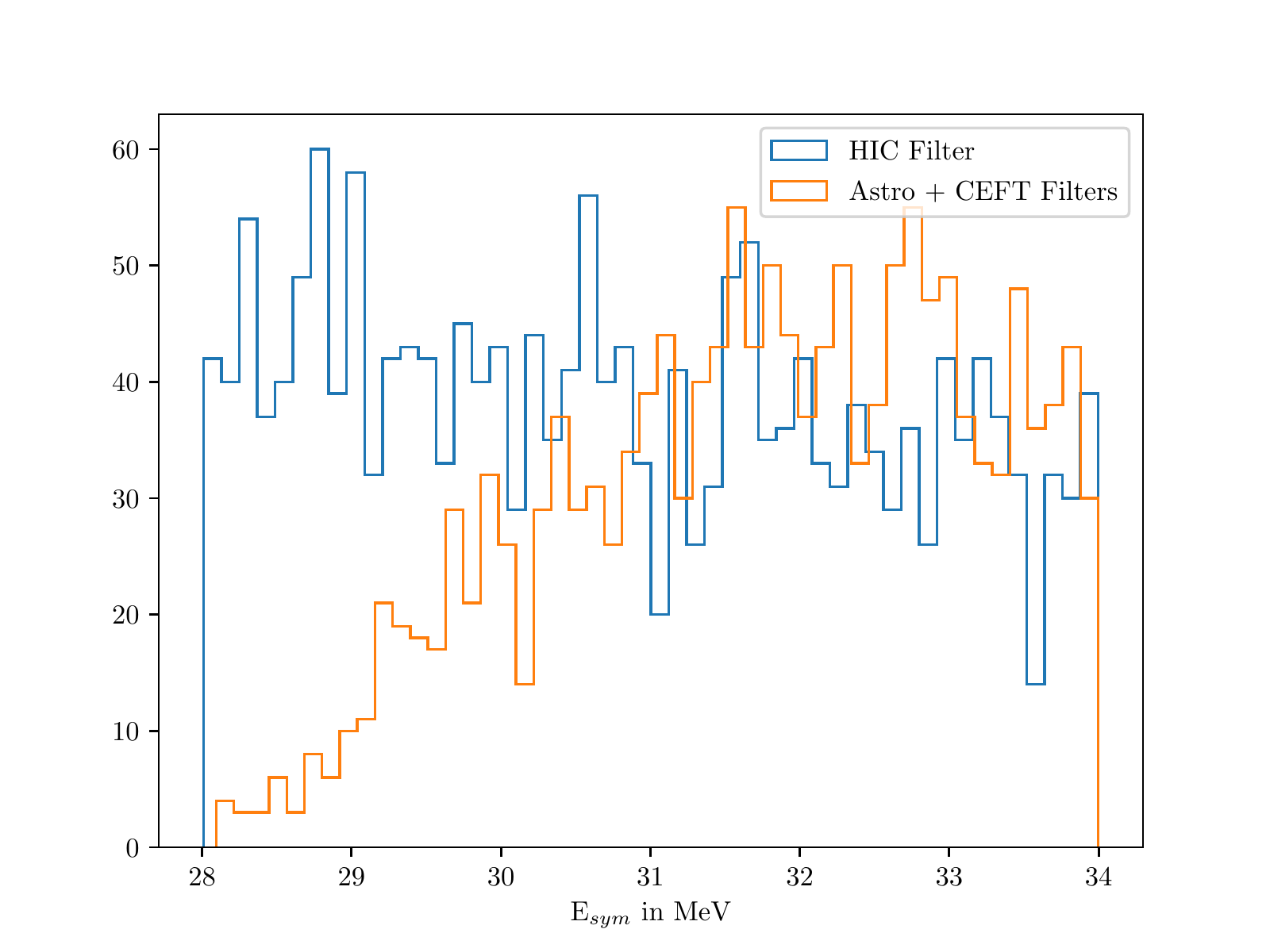}
  \text{\textbf{(B)} Distribution of $E_{sym}$ }
\end{minipage}
\begin{minipage}[c]{0.45\textwidth}
  \centering
  \includegraphics[width=1\linewidth,height=0.78\linewidth, angle=0]{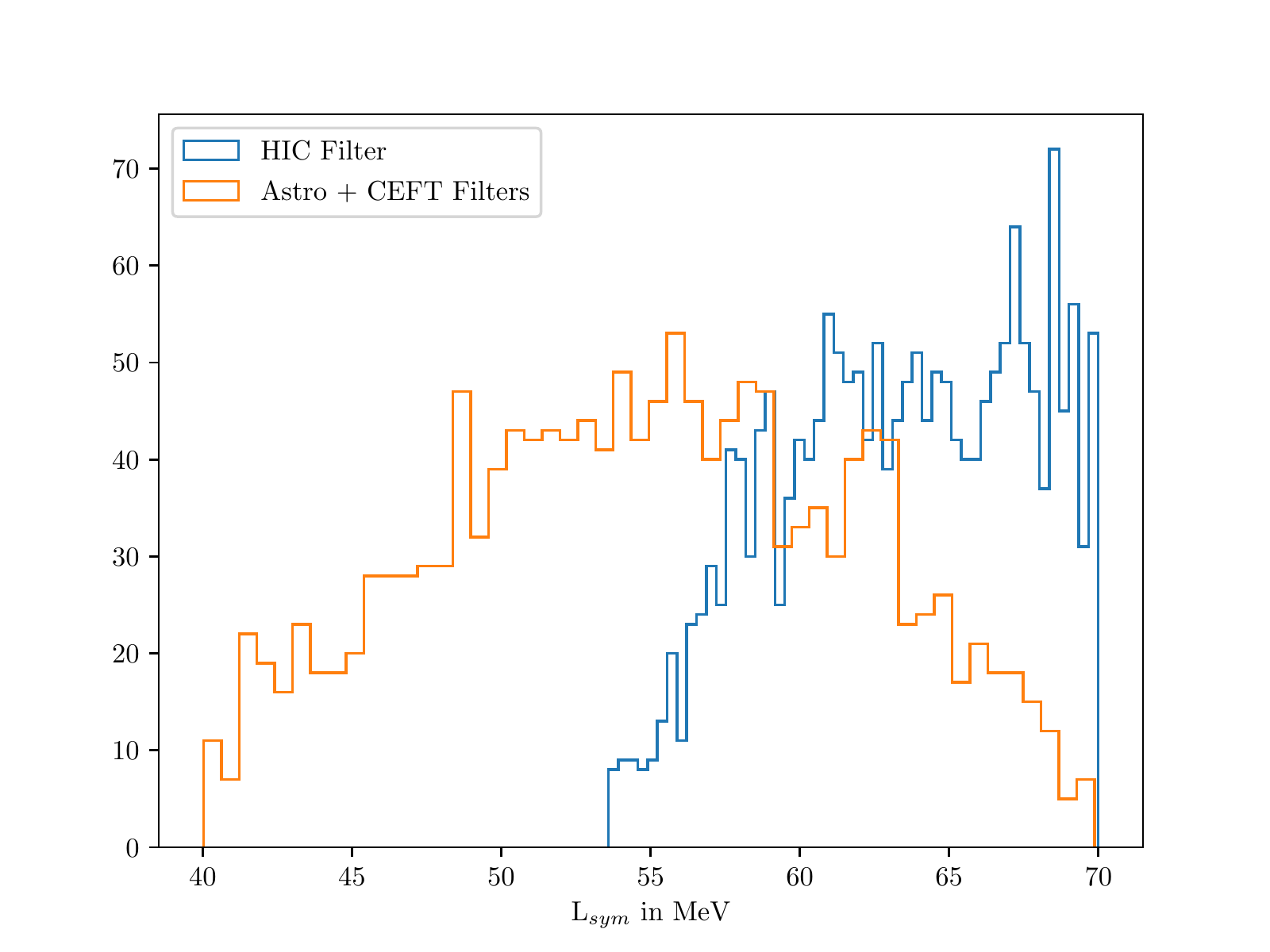}
  \text{\textbf{(C)} Distribution of $L_{sym}$}
\end{minipage}
\begin{minipage}[c]{0.45\textwidth}
  \centering
  \includegraphics[width=1\linewidth,height=0.78\linewidth, angle=0]{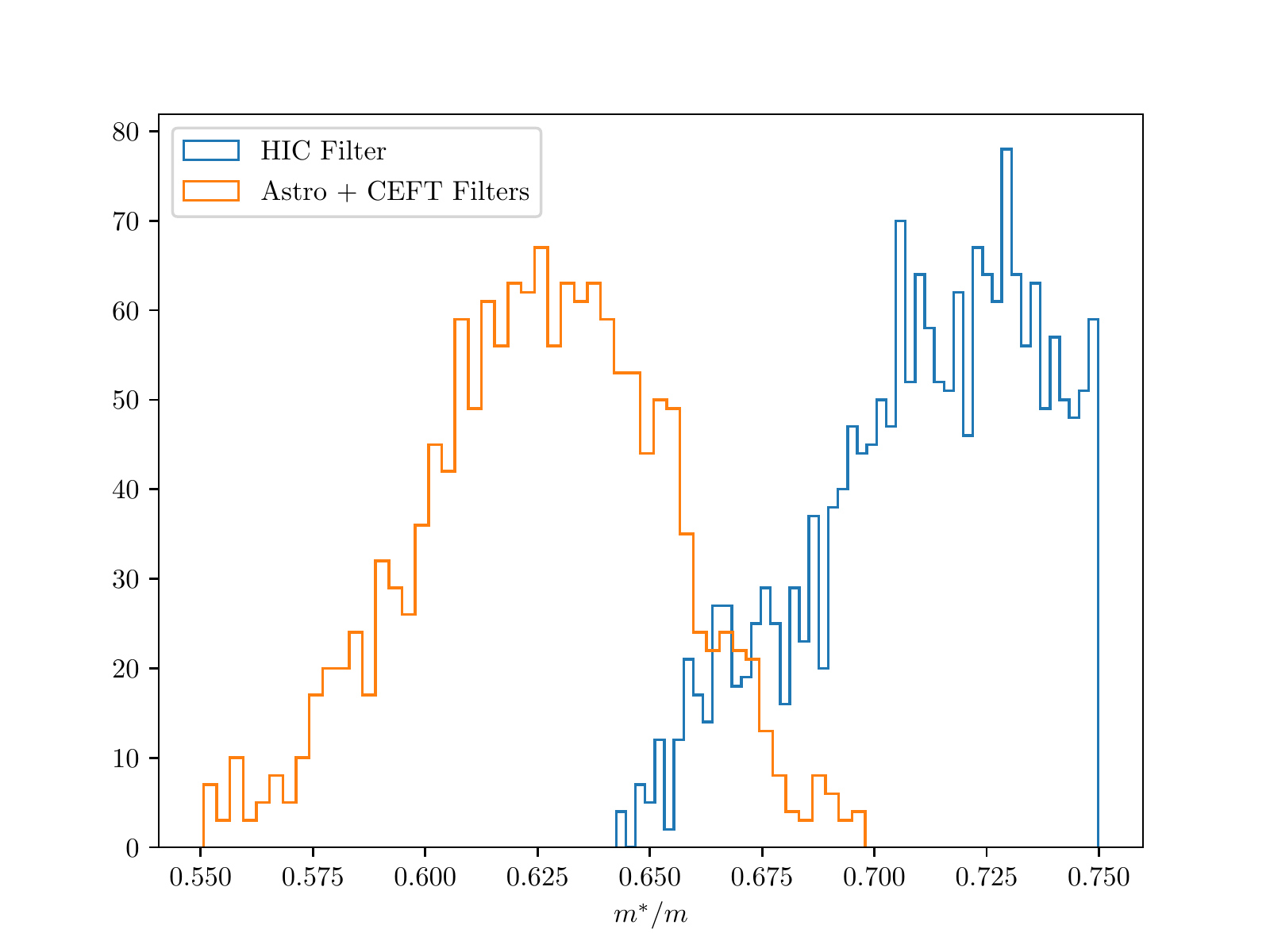}
  \text{\textbf{(D)} Distribution of $m^*/m$ }
\end{minipage}
\caption{Distribution of some nuclear saturation parameters, after application of the HIC filters (KaoS, FOPI, ASY-EOS)}
\label{fig:dist_hic}
\end{figure}

\begin{figure}[htbp]
  \begin{center}
      \resizebox{1\textwidth}{!}{%
{\includegraphics{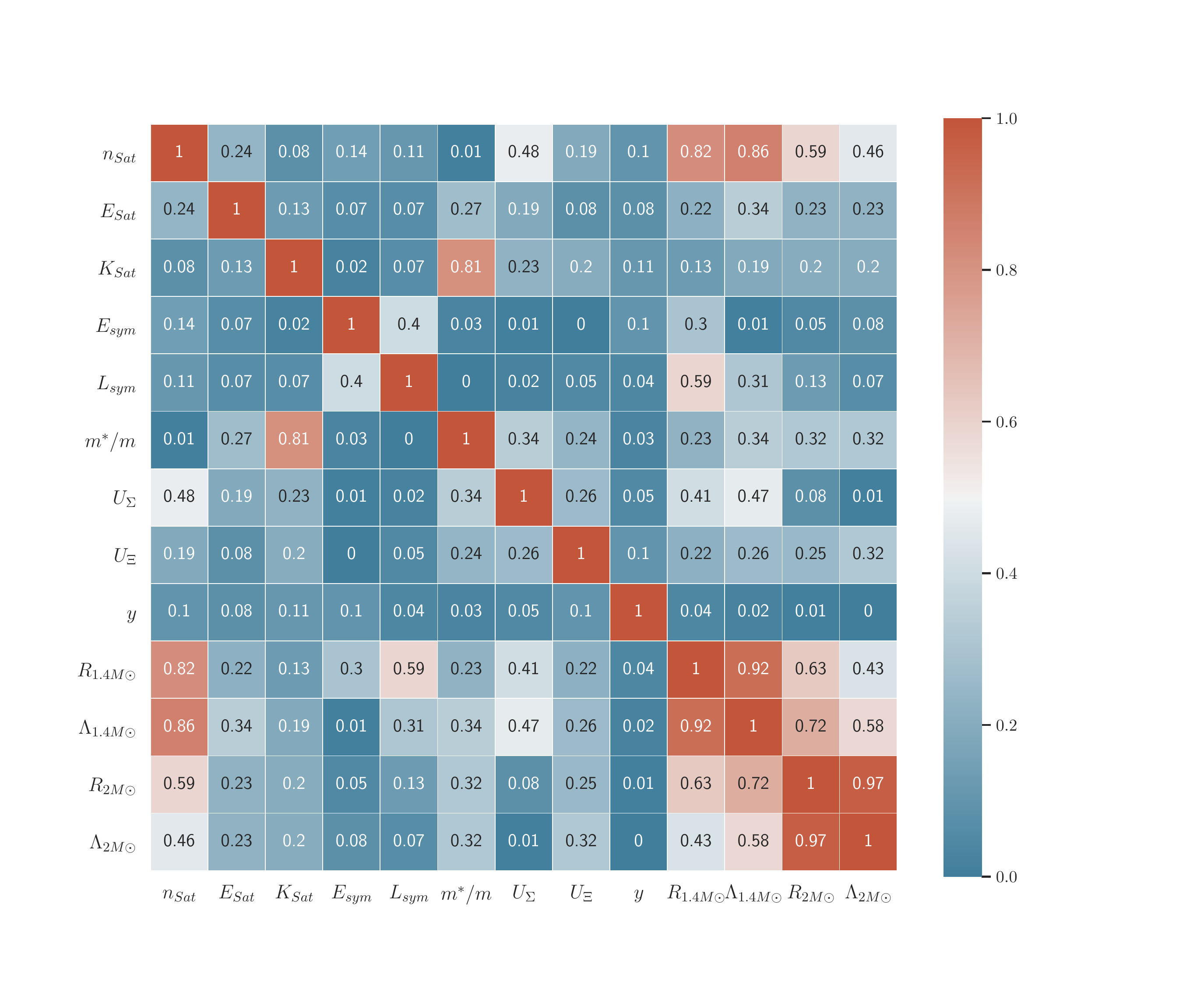}}
}
      \caption{Posterior correlation matrix for variation of nuclear empirical parameters and NS observables, after application of the $\chi$EFT, HIC and NS observations filters} 
    \label{fig:correlations_all} 
  \end{center}
\end{figure}

%%% Please be aware that for original research articles we only permit a combined number of 15 figures and tables, one figure with multiple subfigures will count as only one figure.
%%% Use this if adding the figures directly in the mansucript, if so, please remember to also upload the files when submitting your article
%%% There is no need for adding the file termination, as long as you indicate where the file is saved. In the examples below the files (logo1.eps and logos.eps) are in the Frontiers LaTeX folder
%%% If using *.tif files convert them to .jpg or .png
%%%  NB logo1.eps is required in the path in order to correctly compile front page header %%%

%%% If you are submitting a figure with subfigures please combine these into one image file with part labels integrated.
%%% If you don't add the figures in the LaTeX files, please upload them when submitting the article.
%%% Frontiers will add the figures at the end of the provisional pdf automatically
%%% The use of LaTeX coding to draw Diagrams/Figures/Structures should be avoided. They should be external callouts including graphics.

\end{document}